\RequirePackage{fix-cm}

\documentclass[twocolumn]{svjour3}           % twocolumn

\smartqed  % flush right qed marks, e.g. at end of proof

\usepackage{graphicx}
\graphicspath{{figs/}} % Setting the graphicspath
\usepackage{amsmath} % math
\usepackage{physics} % physics equations
\usepackage[ruled,vlined]{algorithm2e} % pseudo code 2
\usepackage{graphicx}
\usepackage{epsfig}
\usepackage{hyperref}
\usepackage{amssymb}

\usepackage{breakcites} % fix latex's next-line issue
\usepackage{amsfonts}

\begin{document}

\title{Three-dimensional granular flow continuum modeling via material point method with hyperelastic nonlocal granular fluidity
}
% \subtitle{Do you have a subtitle?\\ If so, write it here}
\titlerunning{3D Granular Flow Modeling via MPM-NGF}  % if too long for running head

\author{Amin Haeri \and
        Krzysztof Skonieczny}

\institute{
    A. Haeri \at
        \email{amin.haeri@concordia.ca}
    \and
    K. Skonieczny \at
        \email{krzysztof.skonieczny@concordia.ca}
    \and
    Department of Electrical and Computer Engineering, \\
    Concorida University, Canada
}
% {Amin Haeri}[orcid=0000-0002-1217-656X]
% {Krzysztof Skonieczny}[orcid=0000-0002-6540-3922]

\date{}

\maketitle

\begin{abstract}
    The accurate and efficient modeling of granular flows and their interactions with external bodies is an open research problem. Continuum methods can be used to capture complexities neglected by terramechanics models without the computational expense of discrete element methods. Constitutive models and numerical solvers are the two primary aspects of the continuum methods. The viscoplastic size-dependent nonlocal granular fluidity (NGF) constitutive model has successfully provided a quantitative description of experimental flows in many different conﬁgurations in literature. This research develops a numerical approach, within a hyperelasticity framework, for implementing the dynamical form of NGF in three-dimensional material point method (3D MPM, an appropriate numerical solver for granular flow modeling). This approach is thermodynamically consistent to conserve energy, and the dynamical form includes the nonlocal effect of flow cessation. Excavation data, both quantitative measurements and qualitative visualization, are collected experimentally via our robotic equipment to evaluate the model with respect to the flow geometry as well as interaction forces. The results are further compared with the results from a recent modified plastic Drucker-Prager constitutive model, and in other configurations including wheel-soil interactions, a gravity-driven silo, and Taylor-Couette flow.

\end{abstract}

\section{Introduction}
	Terrestrial construction, mining, and agricultural vehicles are extensively in contact with soil as a granular material. In space, exploration rovers are as well, as will future robots be that will collect and process extraterrestrial soils for in situ resource utilization. However, granular flows and their interactions with rigid/soft bodies are still poorly understood. Besides experiments, simulations can hugely contribute to this end. Nevertheless, the accurate and efficient modeling of such complex flows \cite{Kam17complex} and interactions are still open research problems. One current direction of this research is the discrete element method (DEM), first proposed in \cite{Cun79dem}, which simulates contact mechanics for millions of individual granular particles. But it is so computationally intensive as to be infeasible for online applications in the foreseeable future \cite{comp13}, and for large physical domains can be untenably expensive even in offline applications \cite{DunDemExpK}. DEM's computationally efficient variants such as position based dynamics (PBD) \cite{Mac14flex,Hol14p2} and projective dynamics \cite{Bou14pd} suffer from inaccuracy when their efficiency results from decreasing the degrees of freedom (e.g. number of particles) of the system. On the other hand, several researchers today highlight the insufficient predictive power of classical terramechanics models for arbitrary configurations \cite{comp14n1,comp15new,comp14n2}. Aside from DEM and classical terramechanics, another major direction is using methods from continuum mechanics \cite{God14cm} which compromise between computational efficiency and accuracy. In this case, constitutive models and numerical solvers play important roles in arriving at this compromise.
	
	\begin{figure*}[!t]
        \centering
        \includegraphics[width=1\textwidth]{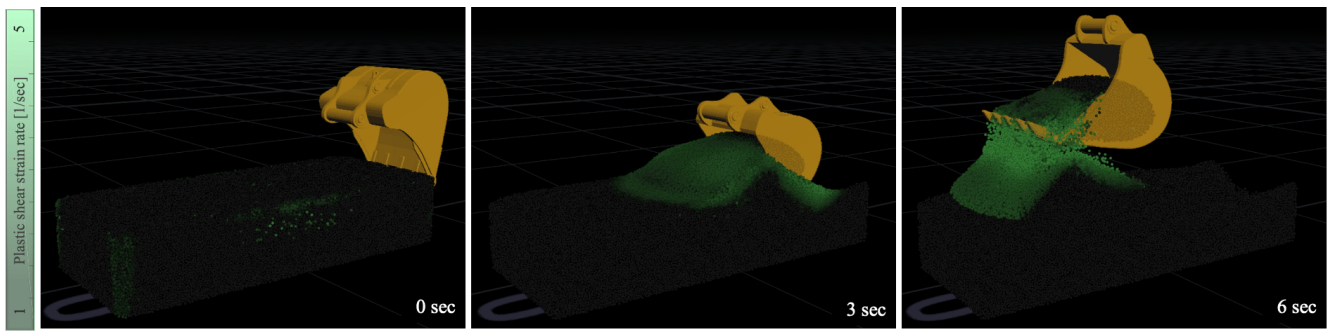}
        \caption{Industrial excavator modeled via MPM-NGF.}
        \label{fig:bucket}
    \end{figure*}
    
	% Numerical solvers
	In continuum mechanics, analytical methods are sometimes proposed as efficient alternatives to numerical methods \cite{am}. However, they might be inaccurate depending on the model simplification assumptions. These assumptions can even make the model inflexible to different conditions and configurations. Among the several continuum-based numerical methods to solve the governing equations of granular flows, finite difference method (FDM), for fluid-like behavior, and finite element method (FEM), for solid-like behavior, are the common base methods. Both methods can yield good results in certain cases. However, FDM has difficulties with extensional disconnection and static regimes, while FEM has issues when mesh distortion becomes large \cite{DunMScPaper}.
	
	% Explanation
	The material point method (MPM) is a modern approach that combines the advantages of both FDM and FEM without having their disadvantages. In other words, MPM is similar to the FEM, but also incorporates the complementary advantage of FDM by keeping an undeformed Cartesian background grid appropriate for large-deformation problems as well \cite{comp12}. In addition to the grid, MPM consists of particles which carry information (mass and momentum) during the simulation. The particles can freely deform and, at each time step, transfer the information to the grid nodes and vice versa. The governing equations can be solved on the active grid nodes (the ones that have mass), identical to FEM. The grid, then, transfers the updated information back to the particles while it remains undeformed. Some recent lines of research have proved MPM as one of the most accurate and efficient methods for granular flow modeling \cite{DunMScPaper,DunDemExpK,mpmHu,klarDP,Hae20isarc,Hae20eas}. Furthermore, instability and non-conserved angular momentum issues observed in the earlier variants including smoothed particle hydrodynamics (SPH) \cite{Mon92sph}, fluid implicit particle (FLIP) \cite{Bri15flip}, and particle-in-cell (PIC) \cite{Har64pic} are addressed and fixed in MPM by using proper particle-grid transfer schemes \cite{apic,polypic}.
	
    % Constitutive models
    Constitutive models are used to calculate the internal forces. They relate stress to strain through the stress tensor in the conservation of momentum equation. Better models can advance the knowledge of how to represent the complex systems of physical interactions with reduced-dimensionality representations. For granular materials, they should generally cover quasi-static elastic (solid-like), intermediate (visco)plastic (fluid-like), and inertial disconnected (gas-like) regimes \cite{Kam10regime}. For the two extreme regimes, elastic and disconnected, models have been developed based on soil mechanics \cite{comp16,comp17,comp20,am} and the kinetic theory of gases \cite{comp15}, respectively. However, the middle (visco)plastic regime is more challenging where the granular material flows somewhat like a liquid. For many years, for this regime, plastic models have been developed by combining various yield criteria and plastic flow rules \cite{DP,naflowRule,klarDP}. In contrast, recently, from fluid mechanics literature, viscoplastic models have made progress by invoking a fluid-like flow approach with an appropriate yield criterion \cite{local,3dStNonK}.
	
    % Explanation
    For the elastic regime, Cauchy elasticity, hyperelasticity and hypoelasticity are the general techniques used and can be considered as a framework for the development of (visco)plasticity techniques. In Cauchy elasticity, the current state of stress depends only on the current state of deformation, and it must be combined with a yield criterion (e.g. Mohr-Coulomb, Drucker-Prager, etc.). While this is the most common framework for granular materials (especially for soils) \cite{soilp}, one issue with this linear model is that it may violate the energy equation of thermodynamics. In hyperelasticity, however, the nonlinear constitutive model is based on the existence of a strain energy function where thermodynamics laws are always satisfied. This framework keeps track of the deformation gradient and is widely used in MPM \cite{mpmCourse}. In hypoelasticity, on the other hand, the stress is a function of the current state of stresses and strains as well as their rates. Thus, hypoelastic models consider reversibility but they typically can not be formulated in a thermodynamically consistent manner \cite{dunMScK}. For the free regime, while models based on kinetic theory for gases can be used, it is also possible to define some saturation density functions to make granular particles free of stress \cite{DunMScPaper}.

    For the plastic regime, incremental theory and/or the deformation theory can be used \cite{soilp}. In deformation theory, the relationship is between the plastic strain and the stress, while in incremental theory it is between the plastic strain rate and the stress and stress rate. Both theories assume the soil is elastic until the stress satisfies a yield criterion. Within the incremental theory, materials are divided into elastic-perfectly plastic and strain hardening ones. While the yield criterion in the former is a function of stress only, in the latter it is a function of plastic strain as well. The flow theory in plasticity defines plastic strain in addition to elastic strain as $d\gamma = d\gamma^\text{e} + d\gamma^\text{p}$. For the plastic part, flow theory provides a plastic flow rule relating the current stress and plastic strain and their incremental changes. This relation is generally assumed to be homogeneous and linear in the incremental changes of stress and plastic strain. This assumption ignores viscosity effects and makes the behavior time-independent \cite{soilp}. In this regime, however, granular materials have similarities with viscoplastic fluids such as Bingham and Herschel-Bulkley fluids \cite{local} which are time- (or rate-) dependent as
    \begin{equation}
        \dfrac{d\gamma}{dt} = \dfrac{d\gamma^\text{e}}{dt} + \dfrac{d\gamma^\text{p}}{dt}  = \dot{\gamma}^\text{e} + \dot{\gamma}^\text{p}.
        \label{equ:vstrain}
    \end{equation}
    
	Plastic models can suffer from rate-independency \cite{soilp} and in some cases, they may have issues with modeling strain hardening \cite{klarDP}. Viscoplastic models eliminate some numerical difficulties associated with plastic models, such as hardening \cite{soilp2}. Among viscoplastic constitutive models, a focus of this work, local models (e.g. \cite{local}) lack robustness in their ability to predict all flow phenomena, while nonlocal models are accepted as highly predictive in different flow regimes \cite{nonlocal}. Nonlocality brings the grain size effect \cite{Kam18size}, secondary rheology (where flow induces flow), and flow cessation (where flow may stop even though the stress appears to be above the yield limit) into the granular flows. Based on three criteria, discussed in \cite{nonlocal}, including ease of implementation, physical motivation, and predictive capability, the nonlocal granular fluidity (NGF) model \cite{2dStNonK,3dStNonK,UnsNonK} outweighs the other nonlocal models, including partial fluidization theory \cite{Ara01pft,Ara02pft,Vol03pft}, I-gradient model \cite{Bou13igm,Bou15igm}, and self-activation model \cite{Pou09sam}.
	
	% Contribution
	The hyperelastic three-dimensional FEM \cite{hypFemK} and hypoelastic two-dimensional MPM \cite{dunMScK,dunPhdK} implementations of the viscoplastic NGF model have already provided promising results in various flow configurations. The additions and contribution of this paper are the development of a hyperelastic numerical approach of the dynamical form of this model and its implementation in three-dimensional MPM. The hyperelasticity and dynamical form ensure the conservation of energy (i.e. thermodynamic consistency) and the occurrence of flow cessation, respectively. The approach is verified by our own experiments as well as some additional test cases found in literature.

    % Outline
    This work is organized in the following manner: \S \ref{sec:theory} will first present a background on the nonlocal theory. Next, \S \ref{sec:imp} will develop a hyperelastic numerical approach of the dynamical nonlocal granular fluidity (NGF) for material point method (MPM). Then, the approach will be validated by multiple experimental and numerical granular cases with large and small elastic strains in \S \ref{sec:verification}.

\section{Theory} \label{sec:theory}
    A constitutive model known as the local model, introduced by \cite{local}, captures some characteristics of dense granular flows - including yield criterion and shear rate dependency - by an analogy to viscoplastic fluids (e.g. Bingham fluids). The local model proposes a relation for the friction coefficient as a function of the inertial number, $I$, for fast ($10^{-3} \lesssim I \lesssim 10^{-1}$) and dense granular flows, given by
    \begin{equation}
        \mu = \mu(I) = \mu_\text{s} + \dfrac{\mu_2-\mu_\text{s}}{I_0/I + 1}
        \label{equ:local}
    \end{equation}
    where $\mu_\text{s}$ is the static (minimum) friction coefficient that causes flow to happen, $\mu_2$ and $I_0$ are the friction coefficient saturation value and a material constant to describe the material rheology, respectively. The inertial number, $I$, is given by
    \begin{equation}
        I = \dot{\gamma}^\text{p} \sqrt{ \dfrac{\rho_\text{s} d^2} {p} }
        \label{equ:inertial}
    \end{equation}
    where $\dot{\gamma}^\text{p}$, $d$, $p$, and $\rho_\text{s}$ are the equivalent plastic shear strain rate, grain diameter, mean normal stress and grain density, respectively. Note that the volume fraction $\phi$ depends linearly on the inertial number $\phi=\phi(I)$ \cite{cteVF}, but can be assumed to be constant for a large range of low $I$. This is a simplifying assumption for well-developed Lagrangian steady-state flows \cite{3dStNonK}.

    The inertial number can also be computed by the inverse of (\ref{equ:local}) and the linearization around $\mu_\text{s}$ \cite{dunMScK} given by
    \begin{equation}
        I = I_0 \dfrac{\mu - \mu_\text{s}}{\mu_2 - \mu_\text{s}}.
        \label{equ:inertial2}
    \end{equation}

    Using (\ref{equ:inertial}) and (\ref{equ:inertial2}) gives the local equivalent plastic shear strain rate
    \begin{equation}
        \dot{\gamma}^\text{p} = {\dot{\gamma}^\text{p}}_{\text{loc}}(p,\mu) =
            \begin{cases}
                \sqrt{p / (\rho_\text{s} d^2)} (\mu-\mu_\text{s}) / b, & \mu \geq \mu_\text{s}\\
                0, & \mu<\mu_\text{s}.
            \end{cases}
        \label{equ:gloc}
    \end{equation}

    This relation will be used in the NGF model presented in the following, where $b=(\mu_2-\mu_\text{s})/I_0$. Also, it shows that the yield criterion for having flow in the local model takes the form of Drucker-Prager criterion. 

    % NGF Model
    The local model works well in the intermediate regime ($\mu \geq \mu_\text{s}$); however, it captures no flow in the quasi-static regime ($\mu<\mu_\text{s}$), and has problems for slow ($I \lesssim 10^{-3}$) flows \cite{3dStNonK}. To capture flows in the quasi-static regime, a nonlocal granular rheology is proposed in \cite{UnsNonK} consistent with modern continuum thermodynamics in an elasto-viscoplastic context. In contrast to the previous constitutive models, in this model the stress depends on the strain and the gradient of strain as well. To consider the effect of nonlocality, this introduces a new scalar variable, $g$, called granular fluidity and is governed by the partial differential equation shown in (\ref{equ:unsNonK}). The granular fluidity ($g$) relates the equivalent plastic shear strain rate ($\dot{\gamma^\text{p}}$) to the friction coefficient ($\mu$) via $g=\dot{\gamma^\text{p}}/\mu$.
    \begin{equation}
        t_0\dot{g} = A^2 d^2 \nabla^2 g - (\mu_\text{s}-\mu)g
                                        - b \sqrt{\dfrac{\rho_\text{s} d^2}{p}} \mu g^2
        \label{equ:unsNonK}
    \end{equation}
    where $\nabla^2 g = \sum_{i=1}^{n}\pdv[2]{g}{\boldsymbol{x}_i}$ and for a constant time-scale $t_0$, a dimensionless material parameter called nonlocal amplitude $A$, the time domain $t$, the i-th component of the spatial domain $\boldsymbol{x}_i$ and the dimension of the spatial domain $n$.

    This model can also be used in a steady-state form, proposed in \cite{2dStNonK} and \cite{3dStNonK} and shown in (\ref{equ:stNonK}). It can be considered as an extension of the local model with modifications addressing its deficiencies.
    \begin{equation}
        \nabla^2 g = \dfrac{1}{\xi^2} (g-g_\text{loc})
        \label{equ:stNonK}
    \end{equation}
    where $g_{\text{loc}} = {\dot{\gamma}^\text{p}}_{\text{loc}}/\mu$,
                $\mu = \tau/{p}$ and
    \begin{equation}
        \xi(\mu) = \dfrac{A}{\sqrt{|\mu-\mu_\text{s}|}} d
    \end{equation}
    for shear stress $\tau$, and the cooperativity length $\xi$ which should be calibrated for each flow geometry by $A$. The cooperativity length is for plastic rearrangement, thereby imposing a length scale ($d$, i.e. grain diameter) on the flow.
    % unity matrix $\mathbf{I}$
    
    The microscopic basis of the model is that flow induces flow; i.e., plastic deformations cause stress fluctuations that can induce plastic events in neighboring material. It makes the model work in the quasi-static flow regime in addition to the intermediate regime. An important feature to note is that when $\mu$ is very close to $\mu_\text{s}$, the cooperativity length is very large. This suggests that when a granular material is in the quasi-static regime ($\mu \leq \mu_\text{s}$), but shear loaded, motion will diffuse over a large range through the network of granular contacts. A small perturbation at one location can be transmitted through this network and through force rearrangement of many of its neighbors \cite{3dStNonK}.

    The generic dynamical form of NGF is derived based upon some fundamental assumptions. The elastic strain is assumed to be small. This, as a kinematical restriction, assumes the rotation tensor $\boldsymbol{R}^\text{e}$ is the dominant part of the polar decomposition of the elastic deformation gradient, as opposed to the stretch tensor $\boldsymbol{U}^\text{e}$, $\boldsymbol{F}^\text{e}=\boldsymbol{R}^\text{e}\boldsymbol{U}^\text{e} \approx \boldsymbol{R}^\text{e}$. The deformation gradient is defined by $\boldsymbol{F}=\pdv{\boldsymbol{x}}{\boldsymbol{X}}$ where $\boldsymbol{x}=\boldsymbol{\chi}(\boldsymbol{X},t)$ is the mapping function between the undeformed and deformed states. The position of points in the undeformed and deformed states is denoted by $\boldsymbol{X}$ and $\boldsymbol{x}$, respectively. Furthermore, the plastic flow is assumed to be incompressible and irrotational. It considers the critical state of materials and ignores the transient dilations and compactions.
    % Theoretically, this assumption considers the volume of material elements between the undeformed and intermediate deformed states constant \cite{UnsNonK}. But it does not prevent the total volume change since we update the total deformation not based on this theory but the velocity spatial gradient.

\section{Approach \label{sec:imp}}
    As discussed, in the case where inertial number $I\gtrsim 10^{-3}$, plastic models and FEM have problems in modeling the rate-dependent intermediate flows. Also, the viscoplastic local model has difficulty in modeling rate-independent quasi-static flows where $I\lesssim 10^{-3}$. Therefore, MPM with viscoplastic nonlocal granular fluidity (NGF) model can be a good candidate approach to model the entire flow.
    
    Here, we develop a hyperelastic numerical approach for the implementation of the dynamical NGF in three-dimensional MPM. This approach also includes the rules for modeling material particles in the inertial disconnected regime. The hyperelasticity framework and the finite difference (FD)-based NGF Laplacian term calculation emulate \cite{hypFemK} and \cite{dunPhdK}, respectively. However, to be more accurate and consistent with our MPM particle-grid transfer scheme, we recommend a 27-point FD scheme for the Laplacian term. Additionally, some aspects of the handling of disconnected particles are from \cite{dunPhdK} while we customize them for the hyperelasticity framework. The MPM code package utilized for generating the results is primarily from \cite{mpmHu}, combined in novel ways with additional elements from \cite{apic,spgrid,pdsampler} written in C++ and Python. We implement our approach as an extension within this MPM for modeling granular flows and their interactions with rigid bodies. Our code is publicly available on GitHub \href{https://github.com/haeriamin/MPM-NGF}{here\footnote{\href{https://github.com/haeriamin/MPM-NGF}{github.com/haeriamin/MPM-NGF}}}.
    
    \subsection{Material Point Method} \label{subsec:mpm}
        \textbf{Governing Equations.}
        Among the conservation laws, mass conservation is automatically satisfied in MPM. Also, the symmetric stress tensor and hyperelasticity will satisfy the angular momentum and energy conservations, respectively. The strong-form unsteady conservation of linear momentum is considered as the governing equation $\phi \rho \dot{\boldsymbol{v}} = \text{div}\boldsymbol{T} + \phi \rho \boldsymbol{G}$, where $\dot{\boldsymbol{v}}$ is the material derivative (i.e. $D(.)/Dt=d(.)/dt+\boldsymbol{v}\nabla (.)$) of velocity $\boldsymbol{v}$, $\boldsymbol{T}$ is Cauchy stress tensor, $\phi$ is volume fraction, and $\boldsymbol{G}$ is gravitational acceleration. MPM uses the weak formulation of the governing equation. It can be obtained by multiplying the strong form by the test function $\boldsymbol{q}$ and integrating by parts over the material (undeformed) domain $\Omega$; given by
        \begin{equation}
            \dfrac{1}{\Delta t}\int_{\Omega}{\rho {\Delta \boldsymbol{v}} \boldsymbol{q} \,dV} =  \int_{\Omega}{\rho \boldsymbol{G} \boldsymbol{q} \,dV} - \int_{\Omega}{\boldsymbol{T} \nabla{\boldsymbol{q}} \,dV}
            \label{equ:weakGov}
        \end{equation}

        Here, the traction on the boundary is ignored by assuming zero first-order boundary conditions. In order to discretize the spatial terms in the governing equation, the Moving Least Squares (MLS) shape function is used. It can speed up MPM \cite{mpmHu} by eliminating the need for explicitly calculating the weighting function derivative, shown in equation (\ref{equ:derivShapeFunc}). Furthermore, it is consistent with the APIC (affine particle-in-cell \cite{apic}) particle-grid transfer scheme in a way that MLS-MPM uses the $\pdv{\boldsymbol{v}}{\boldsymbol{x}}$ quantity from APIC, required in the deformation gradient update, to reduce computational complexity.

        By transferring the left hand side of the equation from the continuum domain $\Omega$ to the particle domain $\Omega_p$, and using the MLS shape function $\Phi(\boldsymbol{x})$, the continuous functions $\boldsymbol{v}$ and $\boldsymbol{q}$ are approximated by the grid node information as
        \begin{equation}
            \begin{split}
            \sum_{p} \int_{\Omega_p}{\rho \boldsymbol{v} \boldsymbol{q} \,dV} & \approx \sum_{p,i,j} {m_p \Phi_i(\boldsymbol{x}_p) \Phi_j(\boldsymbol{x}_p) \boldsymbol{v}_j \boldsymbol{q}_i} \\
            & \approx \sum_{p,i} {m_p \Phi_i(\boldsymbol{x}_p) \boldsymbol{v}_i \boldsymbol{q}_i}
            \label{equ:leftparticle2}
            \end{split}
        \end{equation}
        where $i,j$ and $p$ are the nodal and particle indexes, and the MLS shape function is given by
        \begin{equation}
            \Phi_i(\boldsymbol{x}) = \kappa_i(\boldsymbol{x}_p) \boldsymbol{P}^\text{T}(\boldsymbol{x}-\boldsymbol{x}_p) \boldsymbol{M}^{-1}(\boldsymbol{x}_p) \boldsymbol{P}(\boldsymbol{x}_i-\boldsymbol{x}_p)
            \label{equ:mlsShapeFunc}
        \end{equation}
        with $\boldsymbol{P}$ and $\kappa_i$ as the polynomial basis and localized weighting function, respectively, and $\boldsymbol{M}$ defined as
        \begin{equation}
            \boldsymbol{M}(\boldsymbol{x}_p) = \sum_{i \in B_x} \kappa_i(\boldsymbol{x}_p) \boldsymbol{P}(\boldsymbol{x}_i-\boldsymbol{x}_p) \boldsymbol{P}^\text{T}(\boldsymbol{x}_i-\boldsymbol{x}_p)
            \label{equ:M}
        \end{equation}
        where $B_x$ is a set of grid nodes satisfying $\kappa_i(\boldsymbol{x}_p) \neq 0$. Here, the density function was simplified as $\int_{\Omega_p}{\rho \,dV} = m_p$. Mass lumping approximation \cite{mpmCourse} was also used to eliminate $\Phi_j(\boldsymbol{x}_p)$ by summing it over node $j$ where $\sum_{j}\Phi_j(\boldsymbol{x}_p)=1$.
        
        The derivative of the test function is required to be computed in the right hand side of the governing equation. Hence, adopting a linear polynomial basis $\boldsymbol{P}(\boldsymbol{x}-\boldsymbol{x}_p)=[1, \boldsymbol{x}-\boldsymbol{x}_p]$ and quadratic B-splines as the weighting function $\kappa_i=\boldsymbol{N}_i$, will give the test function derivative
        \begin{equation}
            \nabla{\boldsymbol{q}} = \boldsymbol{M}^{-1}_{p} \boldsymbol{N}_i(\boldsymbol{x}_p) (\boldsymbol{x}_i - \boldsymbol{x}_p)
            \label{equ:derivShapeFunc}
        \end{equation}

        Similarly, the right hand side can also be expressed in the particle domain
        \begin{equation}
            \begin{split}
            \int_{\Omega}{\rho \boldsymbol{G} \boldsymbol{q} \,dV} - & \int_{\Omega}{\boldsymbol{T} \nabla{\boldsymbol{q}} \,dV} = \\
            & \sum_p (\int_{\Omega_p}{\rho \boldsymbol{G} \boldsymbol{q} \,dV} - \int_{\Omega_p}{\boldsymbol{T} \nabla{\boldsymbol{q}} \,dV})
            \end{split}
            \label{equ:rightparticle1}
        \end{equation}

        By substituting (\ref{equ:derivShapeFunc}) in (\ref{equ:rightparticle1}) and using the mass lumping, the following is obtained
        \begin{equation}
            \begin{split}
            \sum_p ( & \int_{\Omega_p}{\rho \boldsymbol{G} \boldsymbol{q} \,dV} - \int_{\Omega_p}{\boldsymbol{T} \nabla{\boldsymbol{q}} \,dV}) = \\
            & \sum_{p,i} [m_p \boldsymbol{G} \boldsymbol{q}_i - V_p \boldsymbol{M}^{-1}_{p} \boldsymbol{T}_p \boldsymbol{N}_i(\boldsymbol{x}_p) (\boldsymbol{x}_i-\boldsymbol{x}_p)]
            \end{split}
            \label{equ:rightparticle2}
        \end{equation}
        where $V_p$ is the particle volume, for the quadratic B-splines $ \boldsymbol{M}^{-1}_{p} = 4/\Delta x^2$, and a one-point quadrature rule is used to express stress continuous function $\boldsymbol{T}$ at a particle ($\boldsymbol{T}_p$).
        
        \textbf{Procedure.}
        \begin{figure*}[!t]
            \centering
            \includegraphics[width=1\textwidth]{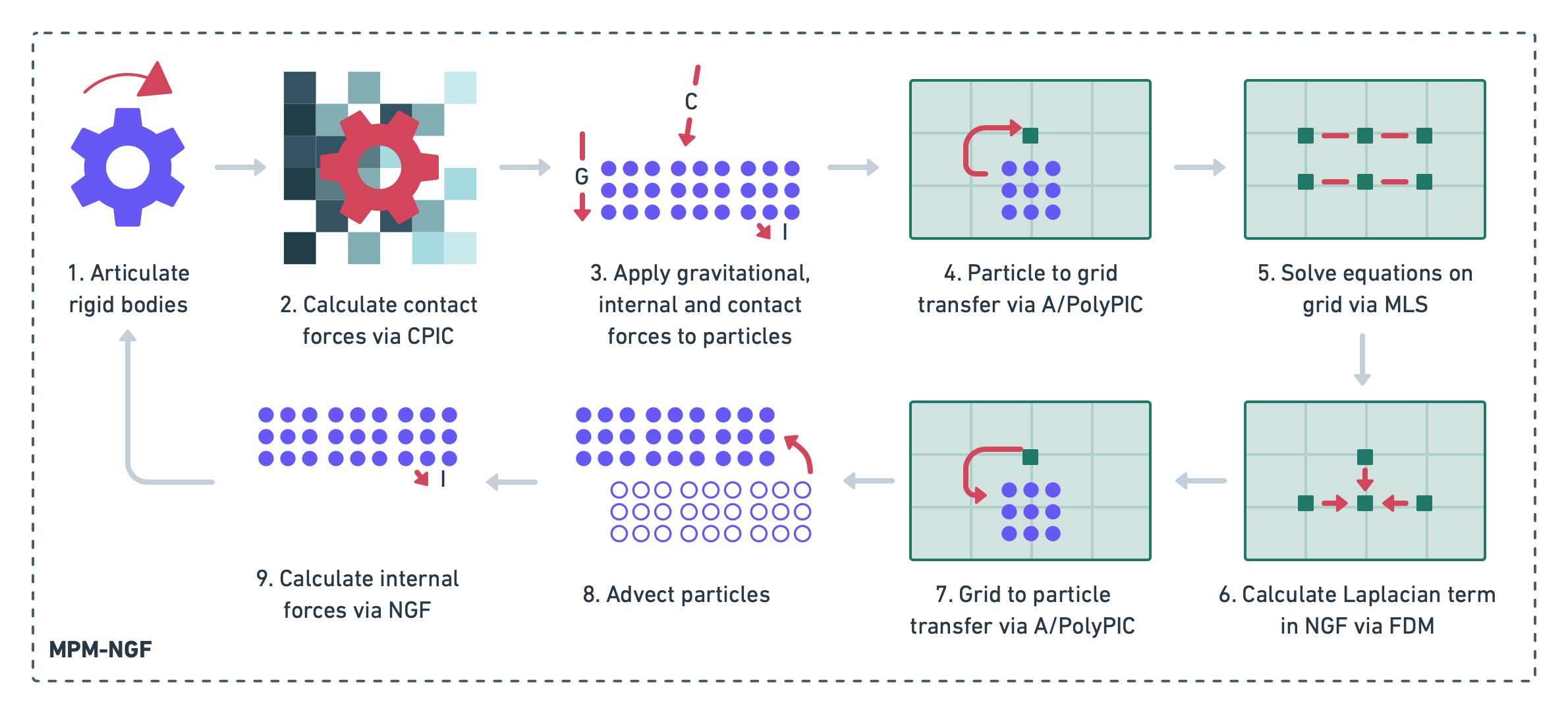}
            \caption{Algorithm overview of material point method with nonlocal granular fluidity model.}
            \label{fig:mpmngfalgo}
        \end{figure*}
        The algorithm in figure \ref{fig:mpmngfalgo} shows the procedure used to generate the results from MPM \cite{mpmHu} with NGF. While Step 6 can be omitted from the MPM procedure, it is introduced to compute the Laplacian term used in NGF model. Also, in Step 9, the calculation of particle internal forces via NGF model is explained in \S \ref{subsec:ngfalgo}. Upon initialization, the 3D background is discretized by an Eulerian grid, the rigid bodies are meshed and the elements get occupied by particles, granular material is discretized by Lagrangian particles. Then, the algorithm enters a loop which repeats every time-step.
        
        First, the rigid bodies are articulated using dynamics equations alongside rigid collisions detection (if any). Then the contacts between rigid bodies and particles of the granular material are detected using CPIC (compatible particle-in-cell) \cite{mpmHu}. This algorithm constructs a narrow-band colored unsigned distance field (CDF) around a rigid body to recognize whether grid nodes in near-boundary particle's support kernel are outside the rigid body or not. With this, two-way particle-rigid body interactions are handled by applying relative impulses on both particles and rigid bodies based on the Coulomb model of friction. This impulse and also the particle internal force (computed in step 9) then contribute to the linear and angular velocity of the rigid body via Euler's equations (rigid body dynamics). Next, gravitational forces, contact forces, plus the internal forces which have been computed in the previous time-step, contribute to the momentum of particles.
        % The mass of a particle is neglected against the mass of the rigid body. This is a valid assumption for our applications where the flow is not highly dynamic.
        
        The background Cartesian grid is essential for accurately solving the momentum equation, calculating spatial derivatives (e.g. velocity spatial gradient and Laplacian term in the NGF model), and detecting collisions. Thus, momentums should be transferred from particles to grid nodes. This is done by APIC (i.e. first-order PolyPIC, also called PolyPIC-4 in 3D) transfer scheme. The particle granular fluidity ($g$) is also directly transferred via a zero-order scheme. This newly introduced scalar variable relates the equivalent plastic shear strain rate ($\dot{\gamma^\text{p}}$) to the friction coefficient ($\mu$) via $g=\dot{\gamma^\text{p}}/\mu$. In contrast to FLIP and PIC, APIC is stable and has smaller dissipation, respectively \cite{apic}. While APIC may suffer from numerical friction (i.e. friction due to the loss in the particle-grid transfer scheme), higher number of scalar modes makes the transfer less dissipative and even lossless \cite{polypic}. Note that the granular fluidity of open-state particles are not transferred \cite{dunPhdK}.

        During solving the momentum equation, the second-order derivative (Laplacian) term in the dynamical NGF can be calculated with the help of background grid and a finite difference (FD) scheme. This technique has also been used in \cite{dunPhdK,stomakhian} as an alternative to the weak form of the finite element method (FEM). In fact, the FD scheme is used to reduce the stencil size and to avoid checkerboard-type modes observed in FEM causing oscillations in the granular fluidity field. This term is obtained using a central 7-point difference scheme as follows
        \begin{equation}
            \begin{split}
            \nabla^2 g_{i,j,k} = & \dfrac{1}{\Delta x^2} (g_{i+1,j,k}+g_{i,j+1,k}+g_{i,j,k+1} - \\
            & 6 g_{i,j,k} + g_{i-1,j,k}+g_{i,j-1,k}+g_{i,j,k-1}).
            \end{split}
            \label{equ:FDLap}
        \end{equation}
        For better consistency with the particle-grid transfer scheme, a central 27-point FD scheme \cite{fd27} can be used instead. It should also be mentioned that, for efficiency, this value is identical for the particles with the same kernel centered at node (i, j, k).

        Next, the boundary conditions can be applied to the boundary grid nodes, and the information on them is transferred back to particles via A/PolyPIC. Here, the gradient of quadratic B-Spline can be approximated by APIC as $\nabla \boldsymbol{N}_{ip} = 4/\Delta x^2 (\boldsymbol{x}_i - \boldsymbol{x}_p) \boldsymbol{N}_{ip}$. This gradient is reused in equation (\ref{equ:derivShapeFunc}). It is also used as the velocity spatial gradient $\boldsymbol{L} = \nabla \boldsymbol{v}_p = \sum_i \boldsymbol{v}_i \nabla \boldsymbol{N}_{ip}$, where this algorithmic optimization reduces the computational load \cite{mpmHu}. Particle positions are then updated based on the updated velocities by $\boldsymbol{x}_p^{n+1} = \boldsymbol{x}_p^n + \boldsymbol{v}_p^{n+1} \Delta t$. Also, dynamic rigid bodies are advected given their scripted motions and/or impulses from particles.

        Now the internal forces between deformable material particles can be obtained via a constitutive model. For granular materials, the constitutive model can be the plastic Drucker-Prager model, the viscoplastic nonlocal granular fluidity (NGF) model, etc. Here, we aim to use the NGF model. It is due to the fact this model is more capable in modeling flows (due to viscoplasticity), and in capturing quasi-static regimes (due to nonlocality) than plastic and viscoplastic local models.
    
    \subsection{NGF Numerical Approach} \label{subsec:ngfalgo}
        \begin{figure}
            \centering
            \includegraphics[width=.5\textwidth]{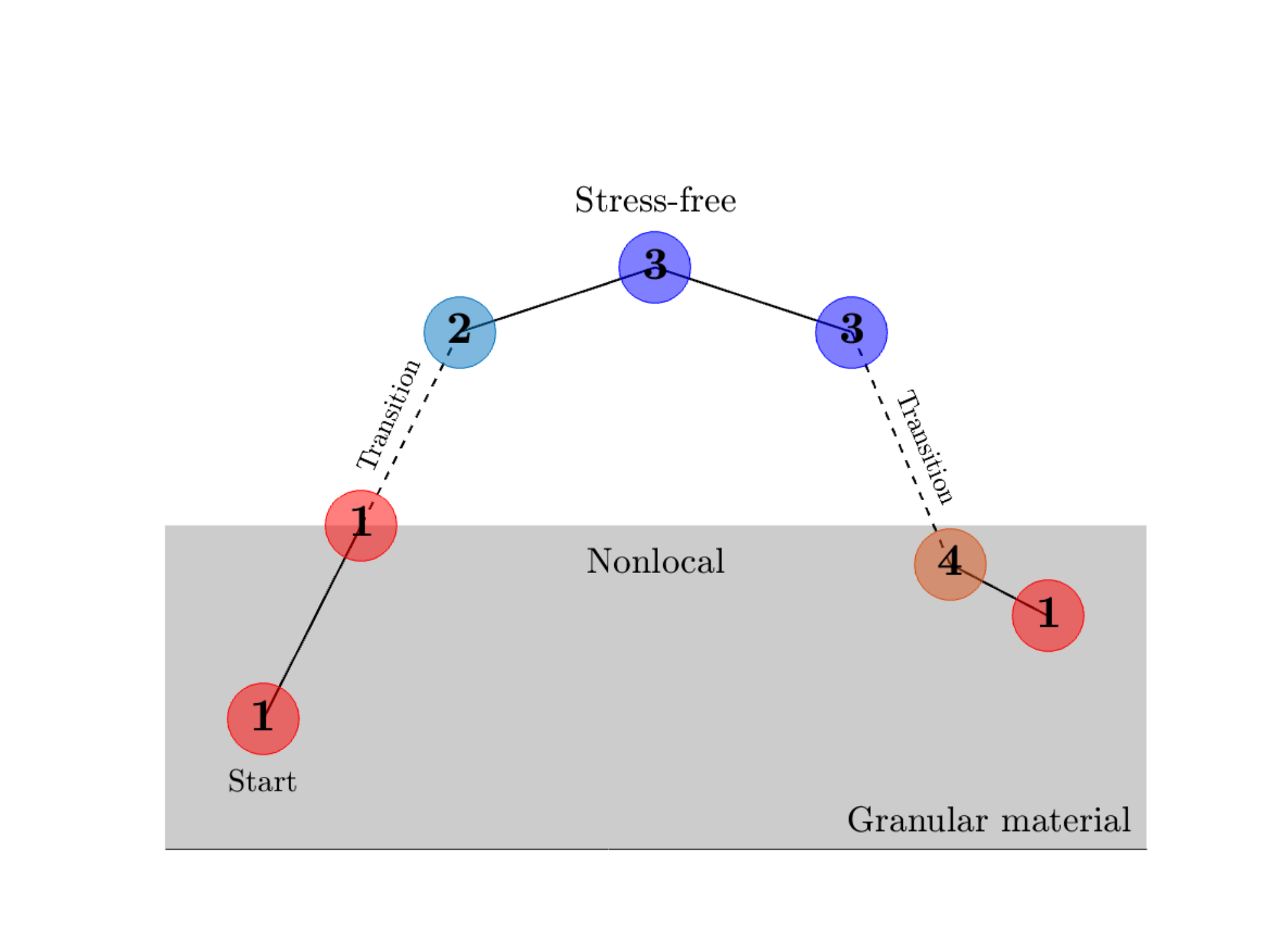}
            \caption{Possible states for a particle. Red (1): under compression, light blue (2): newly stress-free, blue (3): stress-free, light red (4): newly under compression. Gray area represents granular material.}
            \label{fig:states}
        \end{figure}
        
        One novel advance in this research is the accurate calculation of internal forces via the dynamical form of viscoplastic nonlocal granular fluidity (NGF) model \cite{UnsNonK} with hyperelasticity. In fact, this is a thermodynamically consistent version of the nonlocal theory for three-dimensional MPM. Thermodynamic consistency can itself be a goal of using hyperelasticity \cite{HOULSBY2019103167}. In addition, while hyperelasticity is known to be necessary for modeling large elastic strains, some researchers also argue its necessity for \textit{granular flows} with small elastic strains because the relationship between stress and strain is nonlinear in the elastic regime \cite{HOULSBY2019103167,PhysRevE81011303,hutel03199387}. \S \ref{sec:verification} will evaluate granular test cases with both large and small elastic strains. Hyperelasticity requires keeping track of the deformation gradient, which is multiplicatively decomposed into elastic and plastic parts.
        % In the nonlocal theory \cite{hypFemK} the plastic flow is assumed to occur at constant volume so that $det(\boldsymbol{F}^\text{p})=1$.
        The change rate of deformation gradient can be obtained by
        \begin{equation}
            \dot{\boldsymbol{F}}=\boldsymbol{L}\boldsymbol{F}
            \label{equ:dgupdate}
        \end{equation}
        where $\boldsymbol{L}=\pdv{\boldsymbol{v}}{\boldsymbol{x}}$ is approximated by the APIC transfer scheme. Thus, the total deformation gradient is updated by (\ref{equ:dgupdate}), and the elastic deformation gradient is calculated by $\boldsymbol{F}^\text{e}=\boldsymbol{F}(\boldsymbol{F}^\text{p})^{-1}$. The NGF constitutive model is hence utilized to calculate $\boldsymbol{F}^\text{p}$. By assuming that the viscosity ($1/g$) is time-dependent, the PDE of the dynamical NGF model, shown in (\ref{equ:unsNonK}), should be solved for $g$. Then the equivalent plastic shear strain rate can be obtained via $\dot{\gamma^\text{p}}=g \mu$. Note that the friction coefficient $\mu$ is computable using the stress tensor, as shown in the details of (\ref{equ:stNonK}).
        
        Since in MPM, granular materials can be separated, the disconnected particles should also be modeled. While kinetic theory of gases is capable of this modeling, in most cases it is accurate enough to handle granular gas via pure kinematics (stress-free). To detect this regime, pressure (mean normal stress) should be tracked for every individual particle. Figure \ref{fig:states} shows four possible states that can occur for a particle in the next time-step: 1) positive pressure (under compression) 2) first-time negative pressure 3) negative pressure 4) first-time positive pressure. Particles in states 1 and 4 are under elasto-viscoplastic deformations (based on the NGF model); whereas particles with the states 2 and 3 are behaved as stress-free particles. However, as state 1 is distinguishable from state 4 based on prior state, and similarly 2 is from 3, the way particles separate from or return to the material should be modeled. Here, our focus is on the return phase, however, the algorithm is extendable for customizing the separation phase.
            
\begin{figure*}[!t]
            \centering
            \includegraphics[width=1\textwidth]{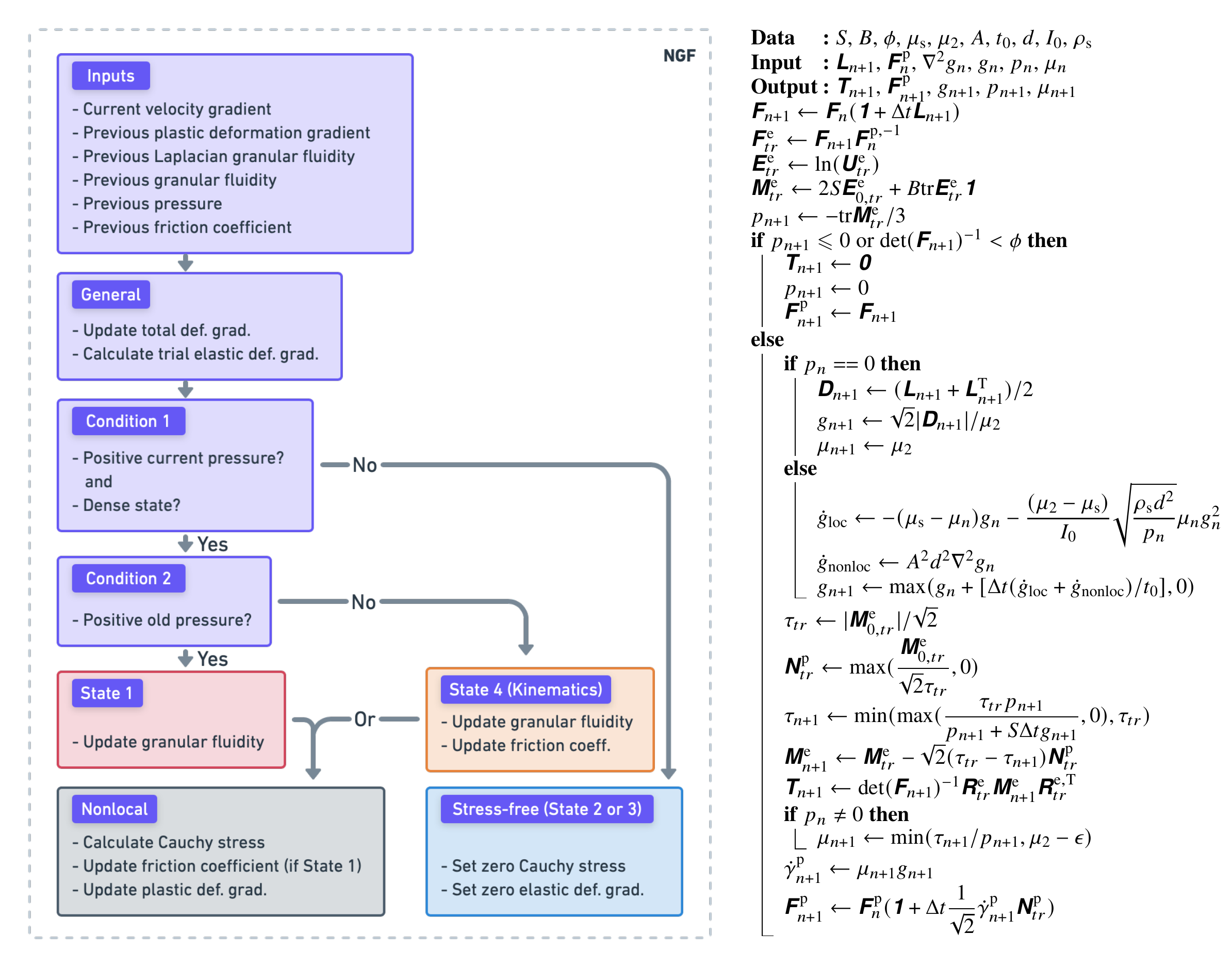}
            \caption{Algorithm overview (left) and pseudo code (right) of nonlocal granular fluidity model for MPM.}
            \label{fig:ngfalgoall}
        \end{figure*}

        As depicted in figure \ref{fig:ngfalgoall}, first, the following steps are taken for particles in any state. The total deformation gradient is updated via the solution from the first-order forward Euler method applied to equation (\ref{equ:dgupdate}) -- while one can use the analytical solution $\text{exp}(\Delta t \boldsymbol{L}) \boldsymbol{F}$ -- as follows ($\boldsymbol{1}$ is unity matrix)
        \begin{equation}
            \boldsymbol{F}_{n+1} = \boldsymbol{F}_{n} (\boldsymbol{1} + \Delta t \boldsymbol{L}_{n+1}).
            \label{equ:dgtotal}
        \end{equation}
    
\begin{figure}
            \centering
            \includegraphics[width=.49\textwidth]{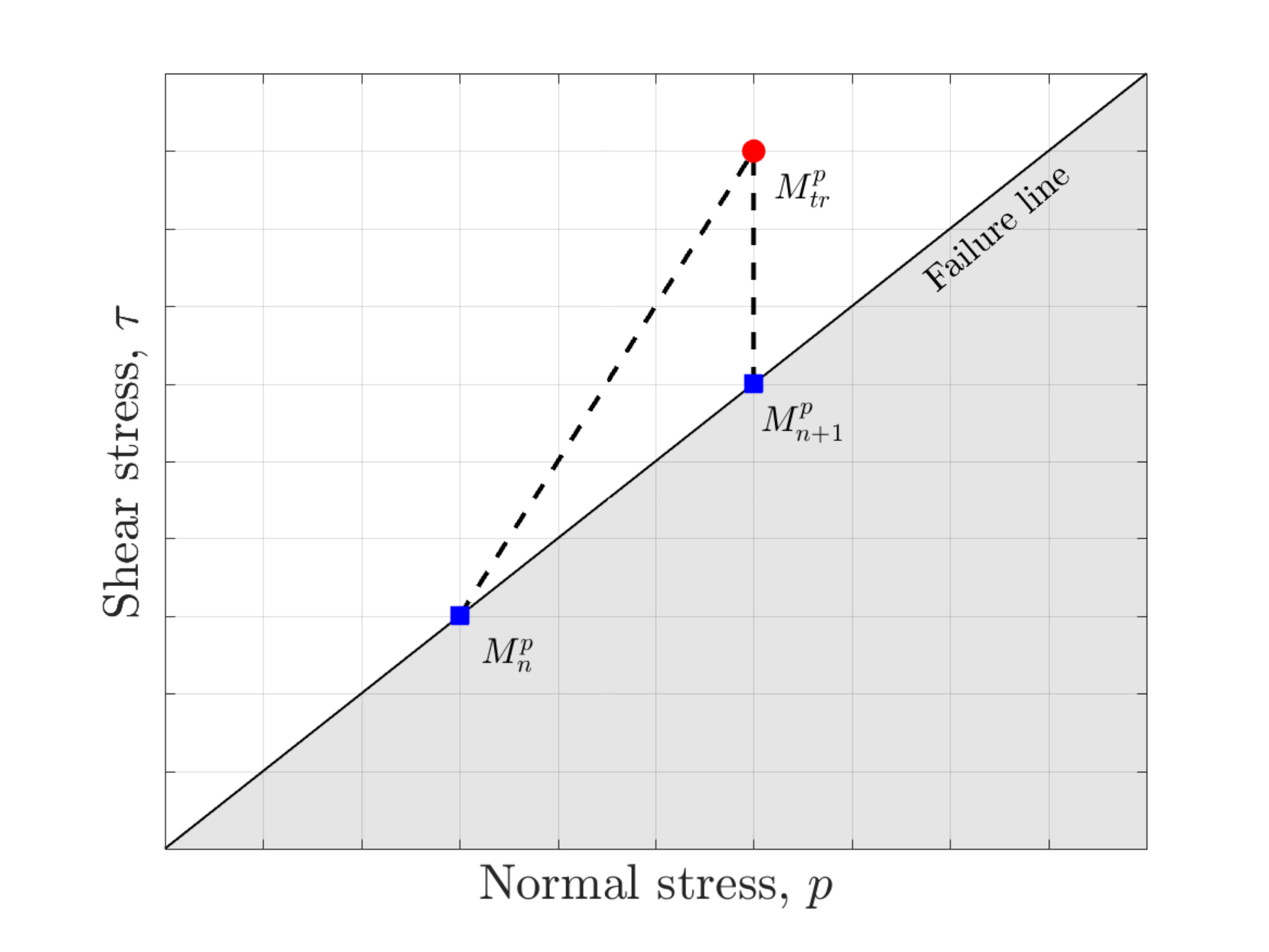}
            \includegraphics[width=.415\textwidth]{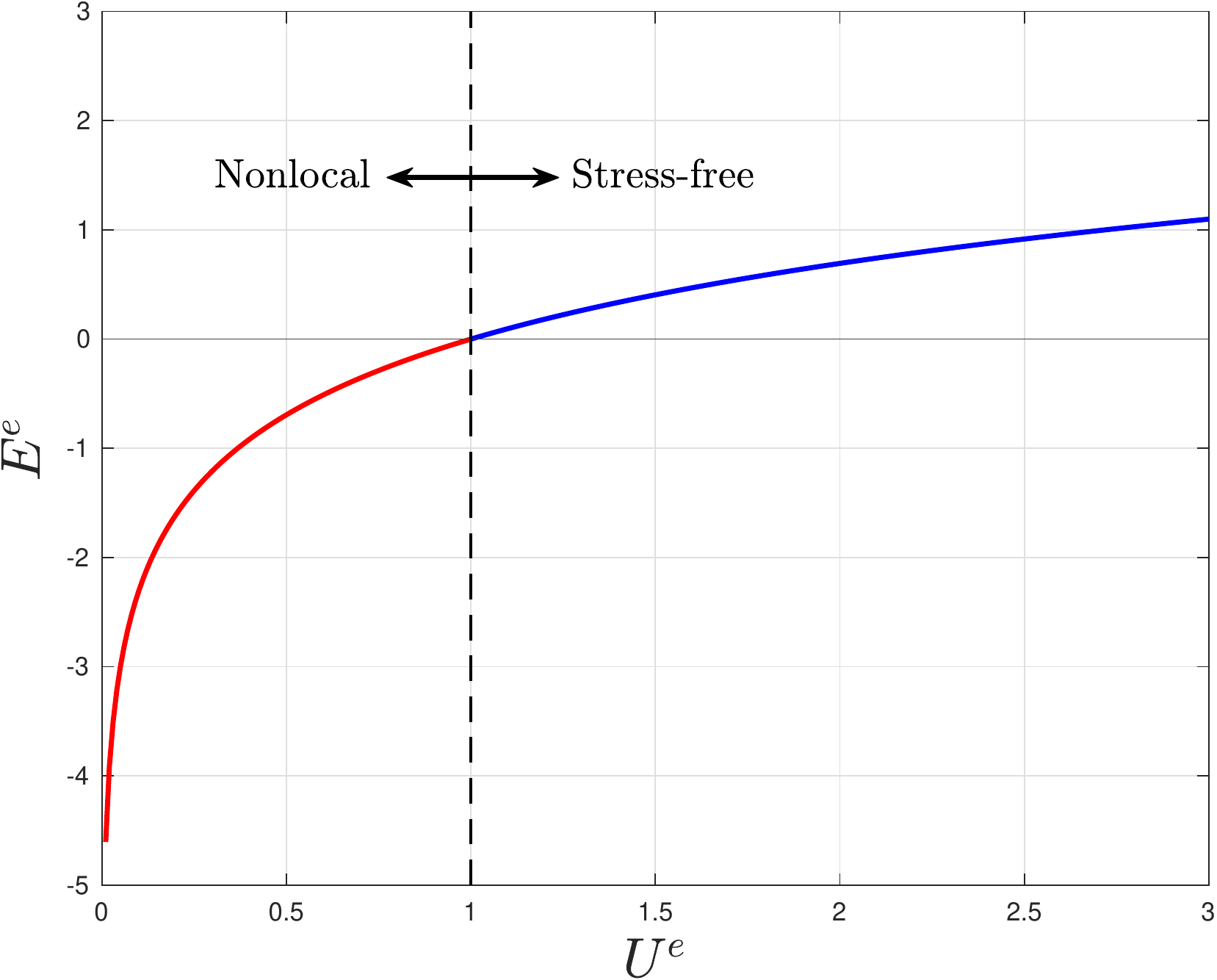}
            \caption{Top: Trial time-step illustration while plastic flow is frozen. Bottom: Particle state illustration with logarithmic Hencky elastic strain in 1D.}
            \label{fig:frozen_hencky}
        \end{figure}

        Now, with the current (i.e. at time-step n+1) total deformation gradient and by calling the previous (i.e. at time-step n) plastic deformation gradient, the elastic deformation gradient can be computed at trial time-step via
        \begin{equation}
            \boldsymbol{F}_{tr}^\text{e} = \boldsymbol{F}_{n+1} \boldsymbol{F}_n^{\text{p},-1}
            \label{equ:dgelasic}
        \end{equation}
        where the trial time-step refers to a middle time-step (i.e. between n and n+1) when the plastic deformation is frozen. Consequently, the trial shear stress differs from the current shear stress as shown in figure \ref{fig:frozen_hencky}-left \cite{Kam15trial}. Now from the right polar decomposition of elastic deformation gradient, one can use the stretch part $\boldsymbol{U}^\text{e}$ to compute the Hencky elastic strain at the trial time-step, given by
        \begin{equation}
            \boldsymbol{E}_{tr}^\text{e} = \text{ln}(\boldsymbol{U}_{tr}^\text{e}).
            \label{equ:hencky}
        \end{equation}
    
        This logarithmic mapping, also shown in figure \ref{fig:frozen_hencky}-right, considers only particles with small elastic strain to be under elasto-viscoplastic deformation. Otherwise, it causes particles with large strain to be stress-free. This is consistent with the assumption made in the nonlocal theory as small elastic strain described before.
    
        As suggested in \cite{hypFemK}, it is convenient to use an elastic Gibbs free energy; Mandel stress $\boldsymbol{M}^\text{e}$ is hence utilized to construct pressure $p$, shear stress $\tau$, direction of plastic flow $\boldsymbol{N}^\text{p}$, and Cauchy stress $\boldsymbol{T}$. Trial Mandel stress is computed by a common constitutive law in non-linear elasticity known as generalized Hookean law \cite{contmechnotes}
        \begin{equation}
            \boldsymbol{M}_{tr}^\text{e} = 2 S \boldsymbol{E}_{0,tr}^\text{e} + B \text{tr}\boldsymbol{E}_{tr}^\text{e} \boldsymbol{1}
            \label{equ:Mtr}
        \end{equation}
        where $E_0$ is the deviatoric part of Hencky strain, and $S$ and $B$ are shear and bulk moduli. The current (and trial) pressure is finally computed by
        \begin{equation}
            p_{n+1} = p_{tr} = -\text{tr}\boldsymbol{M}_{tr}^\text{e}/3.
            \label{equ:pn1}
        \end{equation}
        
        Now depending on the updated pressure $p_{n+1}$ and the density change,
        % \begin{equation}
        %     \dfrac{\rho_{n}}{\rho_{n+1}} \propto J=\text{det}(\boldsymbol{F}_{n+1}),
        %     \label{equ:j}
        % \end{equation}
        the state of the particle is determined. If the pressure is negative or the current density is below the bulk density ($\text{det}(\boldsymbol{F}_{n+1})^{-1}=J^{-1}<\phi$), the particle is in state 2 or 3. Thus, setting zero Cauchy stress tensor makes the particle stress-free and the motion is purely given by kinematics. In addition, as there would be no elastic deformation here, the current plastic deformation gradient is updated via the current total deformation gradient, $\boldsymbol{F}^\text{p}_{n+1} = \boldsymbol{F}_{n+1}$. Note, resetting the current pressure to zero is only for tagging the particle.
        
        Otherwise, it means that the particle is in state 1 or 4. In the case that the previous state of the particle was stress-free ($p_{n} = 0$), the particle is now in state 4; and the granular fluidity should be calculated via kinematics given by
        \begin{equation}
            g_{n+1} = \sqrt{2}|\boldsymbol{D}_{n+1}|/\mu_2
            \label{equ:mngfk}
        \end{equation} % and the deviatoric part $D_0 = D - trD/3  \boldsymbol{1}$
        where an additive decomposition of the velocity spatial gradient $\boldsymbol{L} = \boldsymbol{D} + \boldsymbol{W}$ results in the stretch part $\boldsymbol{D} = (\boldsymbol{L} + \boldsymbol{L}^\text{T})/2$. In addition, the current friction coefficient is set to the friction coefficient saturation value ($\mu_2$) \cite{dunPhdK} based on the $\mu(I)$-rheology \cite{local}.
        
        However, if the previous pressure was non-zero (state 1), the granular fluidity is normally updated using equation (\ref{equ:unsNonK}) and the value of Laplacian term obtained from equation (\ref{equ:FDLap}) as follows
        \begin{equation}
            \begin{split}
            g_{n+1} = g_n + \Delta t [ & A^2 d^2 \nabla^2 g_n - (\mu_\text{s} - \mu_n) g_n - \\
            & \dfrac{(\mu_2 - \mu_\text{s})}{I_0} \sqrt{\dfrac{\rho_\text{s} d^2}{p_n}} \mu_n g_n^2]/t_0.
            \end{split}
            \label{equ:mngf}
        \end{equation}
    
        Then, the remaining associated trial quantities are calculated as follows
        \begin{equation}
        \begin{split}
            &\tau_{tr} = |\boldsymbol{M}_{0,tr}^\text{e}|/\sqrt{2} \\
            &\boldsymbol{N}_{tr}^\text{p} = \dfrac{\boldsymbol{M}_{0,tr}^\text{e}}{\sqrt{2}\tau_{tr}}.
        \end{split}
        \end{equation}
    
        It is assumed that the deviator of Mandel stress and the direction of plastic flow are codirectional. Hence, the plastic flow rule is non-associated and prevents plastic volume change. 
    
        In order to compute the Cauchy stress, it is required to update shear stress via the updated granular fluidity, and then update Mandel stress by
        \begin{equation}
        \begin{split}
            &\tau_{n+1} = \dfrac{\tau_{tr} p_{n+1}}{p_{n+1} + S \Delta t g_{n+1}} \\
            &\boldsymbol{M}_{n+1}^\text{e} = \boldsymbol{M}_{tr}^e - \sqrt{2} (\tau_{tr}-\tau_{n+1}) \boldsymbol{N}_{tr}^\text{p}.
        \end{split}
        \end{equation}
    
        Finally, the Cauchy stress is computed for particles in state 1 or 4 as follows
        \begin{equation}
            \boldsymbol{T}_{n+1} = J^{-1} \boldsymbol{R}_{tr}^\text{e} \boldsymbol{M}_{n+1}^\text{e} \boldsymbol{R}_{tr}^{\text{e},\text{T}}
            \label{equ:CS}
        \end{equation}
        where $\boldsymbol{R}^\text{e}$ is the rotation tensor of the polar decomposition of the elastic deformation gradient. Now with the non-zero previous pressure the current friction coefficient is updated via
        \begin{equation}
            \mu_{n+1} = \tau_{n+1}/p_{n+1},
            \label{equ:mu}
        \end{equation}
        otherwise, it remains $\mu_2$. The equivalent plastic shear strain rate is then calculated via
        \begin{equation}
            \dot{\gamma}_{n+1}^\text{p} = \mu_{n+1} g_{n+1}
            \label{equ:pgamma}
        \end{equation}
        which results in the plastic part of velocity spatial gradient \cite{hypFemK} known as (visco)plastic flow rule given by
        \begin{equation}
            \boldsymbol{L}_{n+1}^\text{p} = \dfrac{1}{\sqrt{2}} \dot{\gamma}_{n+1}^\text{p} \boldsymbol{N}_{tr}^\text{p}.
            \label{equ:pgvel}
        \end{equation}
    
        Therefore, the updated plastic deformation gradient is given by the forward Euler solution of equation (\ref{equ:dgupdate}) as follows
        \begin{equation}
            \boldsymbol{F}_{n+1}^\text{p} = \boldsymbol{F}_n^\text{p} (\boldsymbol{1} + \Delta t \boldsymbol{L}_{n+1}^\text{p}).
            \label{equ:dgplastic}
        \end{equation}
    
        After all, the particle internal force can be calculated via the updated Cauchy stress tensor as $\boldsymbol{f}_p = V_p \ \text{div}\boldsymbol{T}$. There are some extra instructions in the pseudo code shown in figure \ref{fig:ngfalgoall}-right regarding programming limitations which were not included in the aforementioned steps.

\section{Verification and Validation} \label{sec:verification}

    \begin{table*}[!t]
        \begin{center}
        \begin{tabular}{ccccccccc}
        % $S$ [MPa] &$B$ [MPa] &$d$ [mm] &$\phi$ &$t_0$ [s] &$A$ &$I_0$ &$\mu_\text{s}$ &$\mu_2$ \\
        % \hline
        % \{5.8, 1\} &\{12.5, 10\} &\{0.3, 4\} &\{0.67, 0.65\} &1e-4 &0.48 &0.278 &\{0.70, 0.38\} &\{0.96, 0.64\}
        $S$ [MPa] &$B$ [MPa] &$\rho_\text{s}$ [kg/m\textsuperscript{3}] &$d$ [mm] &$t_0$ [s] &$A$ &$I_0$ &$\mu_\text{s}$ &$\mu_2$ \\
        \hline
        \{5.8, 1\} &\{12.5, 10\} &\{2583, 2550\} &\{0.3, 4\} &1e-4 &0.48 &0.278 &\{0.70, 0.38\} &\{0.96, 0.64\}
        \end{tabular}
        \caption{Material parameters for NGF model. First set members are for Excavation, Industrial Excavator, Wheel and 3D Silo (unless otherwise noted), and second ones are for Taylor-Couette Flow.}
        \label{tab:ngfpar}
        \end{center}
    % \end{table} 
        \bigskip
    % \begin{table}
        \begin{center}
        \begin{tabular}{lccccccc}
        Model &$E$ [MPa] &$\Delta t$ [s] &$\Delta x$ [m] &Particle \# (ppc) &Runtime [h] &Runtime/sec [s] &MPE [\%]\\
        \hline
        TC Flow &3.00 &1.0e-4 &4.0e-3 &500K (4) &9.56 &1565 &0.9 \\
        Excavation &15.00 &4.0e-5 &3.3e-3 &230K (8) &4.28 &770 &-0.5 \\
        Excavation* &0.15 &3.5e-4 &4.0e-3 &15K (1) &0.06 &11 &15.8 \\
        Wheel &15.00 &1.0e-5 &3.3e-3 &50K (4) &14.01 &2522 &5.2 \\
        Industrial Excavator &0.15 &1.0e-4 &3.3e-3 &256K (8) &2.17 &1300 &-- \\
        3D Silo &0.15 &1.0e-4 &4.0e-3 &32K (8) &2.87 &857 &--
        \end{tabular}
        \caption{Specifications for simulations of 20-sec 5-cm depth Excavation, 6-sec Industrial Excavator, 20-sec 20\% slip wheel, 12-sec 3D Silo, and 22-sec Taylor-Couette (TC) Flow. Mean percentage errors (MPE) in TC Flow angular velocity, Excavation forward force, and Wheel load are specified relative to experiments. * indicates Excavation model with relaxed Young's modulus for significant computational efficiency yet acceptable accuracy.}
        \label{tab:mpmngfspec}
        \end{center}
    \end{table*}
    
    In this section, we will show the validity and accuracy of our entire MPM-NGF, the nonlocal model itself, and the hyperelasticity framework. We will use experimental \cite{exp1,exp2,exp3} and numerical \cite{hypFemK} Taylor-Couette (TC) Flow results to verify our MPM-NGF and the nonlocal model. Furthermore, we will use our experimental Excavation results to verify our hyperelastic implementation, validate the accuracy of MPM-NGF, and show the necessity of the nonlocal model for capturing a phenomenon in soil-cutting operations. We will also compare the NGF results with the results from a recent modified Drucker-Prager (DP) plasticity model \cite{klarDP} for Excavation. This will show how a viscoplastic model (NGF) can alleviate the hardening modeling as opposed to a plastic model (DP). In addition, three extra test cases including a Wheel, Industrial Excavator, and 3D Silo are modeled and validated experimentally, visually, and theoretically, respectively. In particular, the Wheel case provides a preliminary investigation of MPM-NGF's ability to capture mobility effects; and Industrial Excavator and 3D Silo will depict how MPM-NGF can model the three states the granular material simultaneously experiences in practical cases.
    
    \begin{figure}
        \centering
        \includegraphics[width=.5\textwidth]{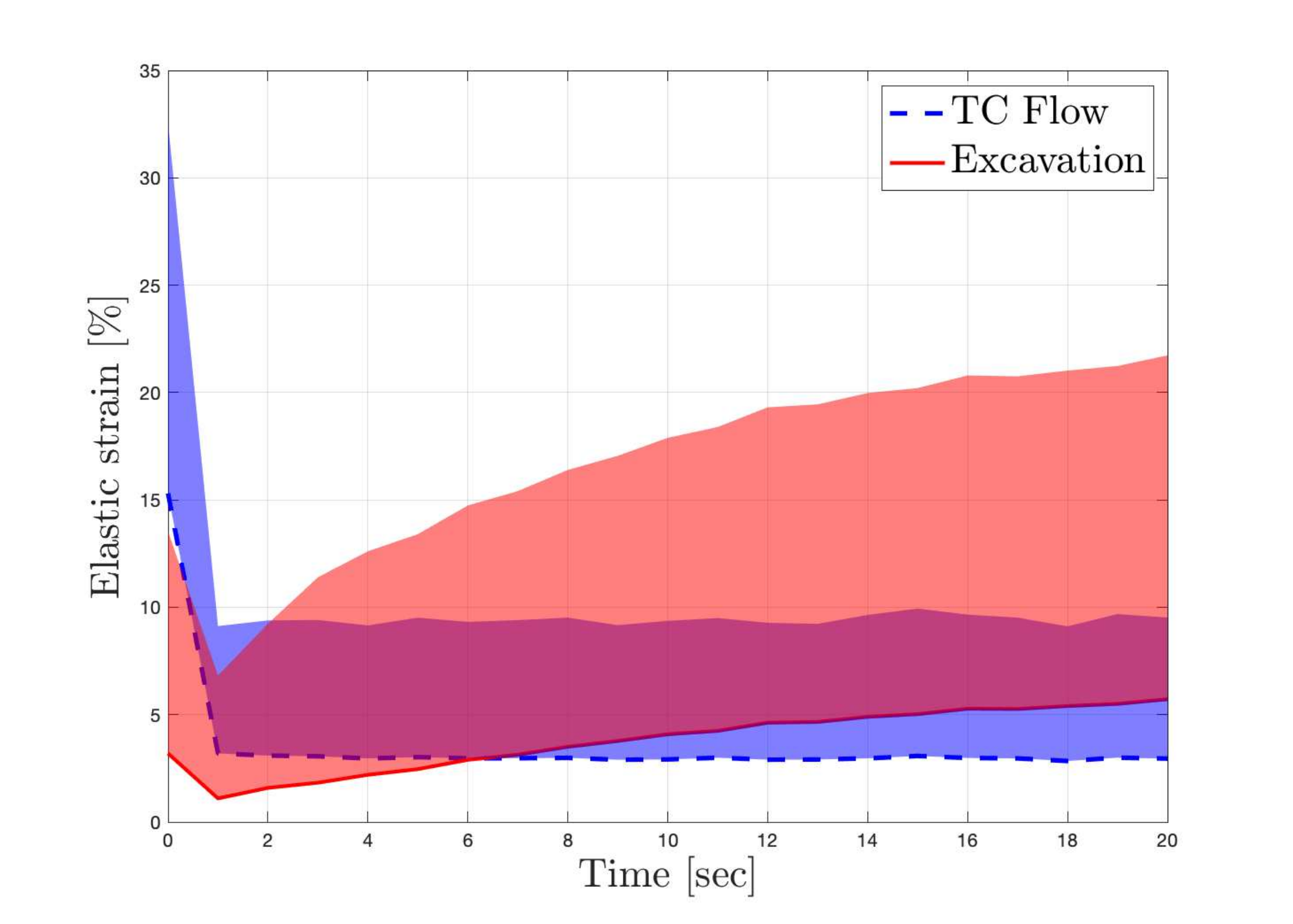}
        \caption{Temporal evolution of mean elastic strain in TC Flow (dashed line) and Excavation (solid line). The colored areas depict one standard deviation above the mean (of elastic strains only).}
        \label{fig:hyper1}
    \end{figure}
    
    \begin{figure}
        \centering
        \includegraphics[width=.5\textwidth]{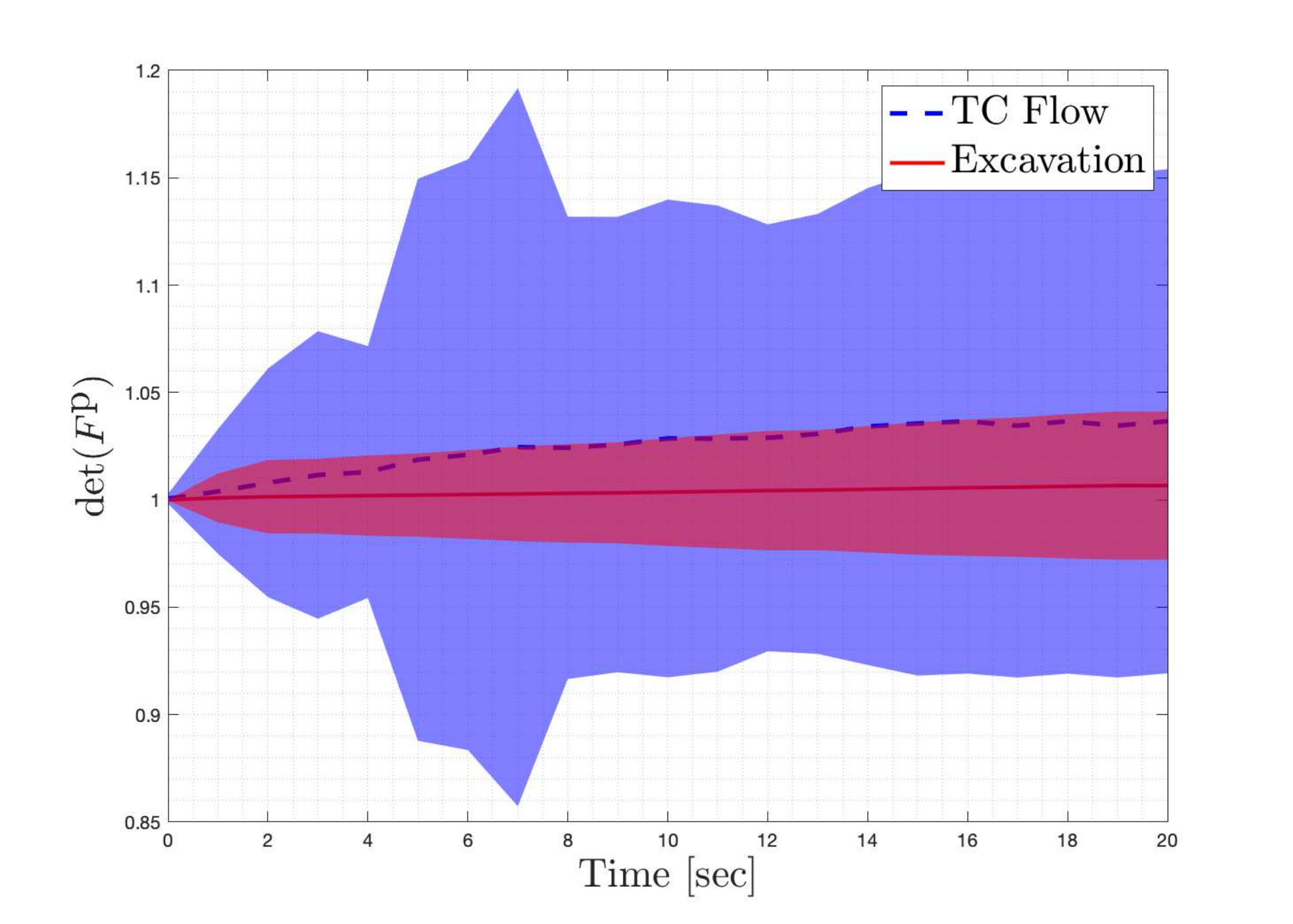}
        \caption{Temporal evolution of mean plastic deformation gradient and its inverse in TC Flow and Excavation. The colored areas are one standard deviation from the mean.}
        \label{fig:hyper2}
    \end{figure}
    
    \begin{figure}
        \centering
        \includegraphics[width=.41\textwidth]{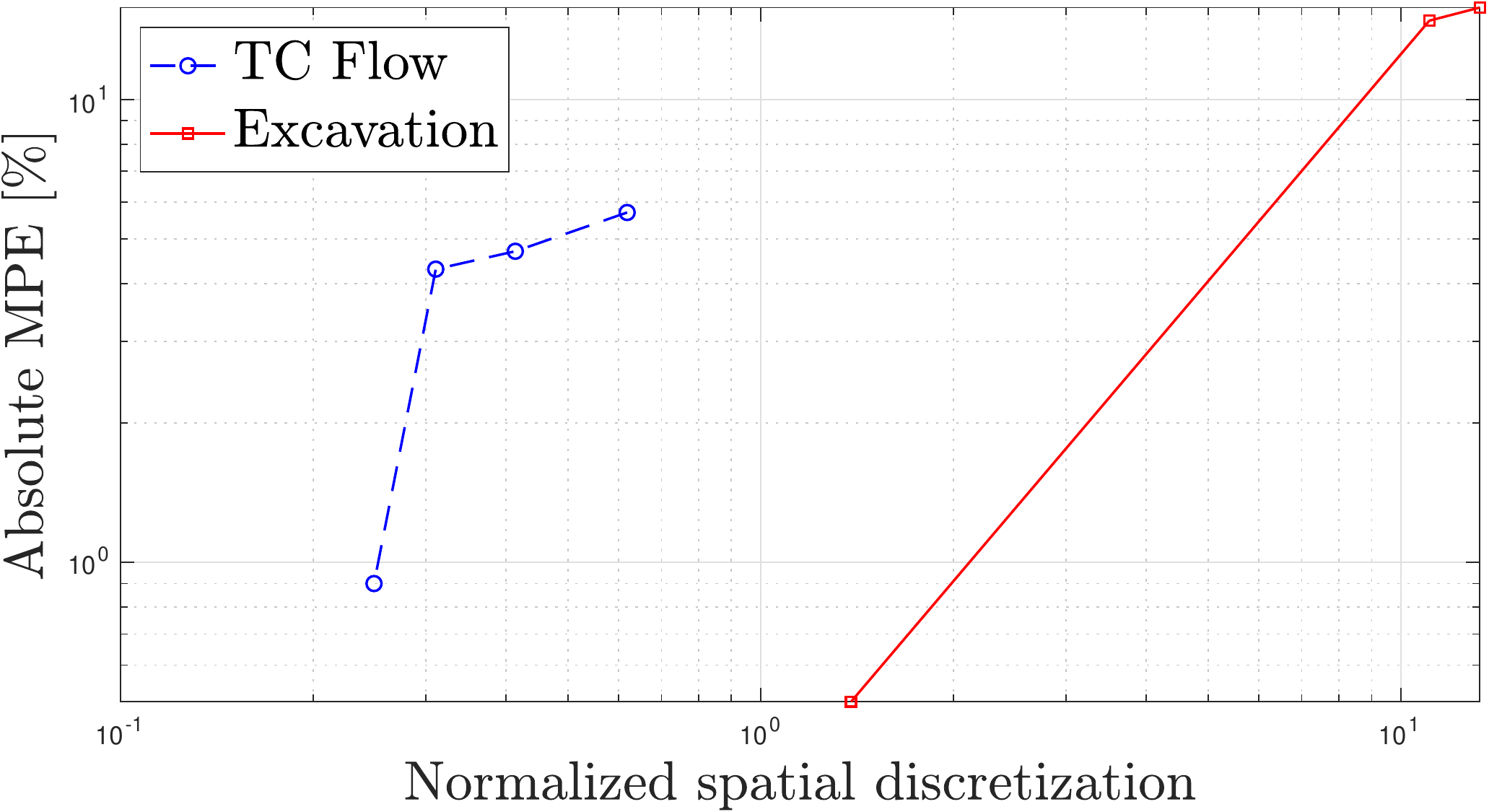}
        \caption{Convergence of absolute mean percentage error (MPE) versus normalized spatial discretization, $\Delta x / (d*\text{ppc})$, in TC Flow and Excavation.}
        \label{fig:meshconv}
    \end{figure}
    
    \begin{figure*}
        \centering
        \includegraphics[width=.38\textwidth]{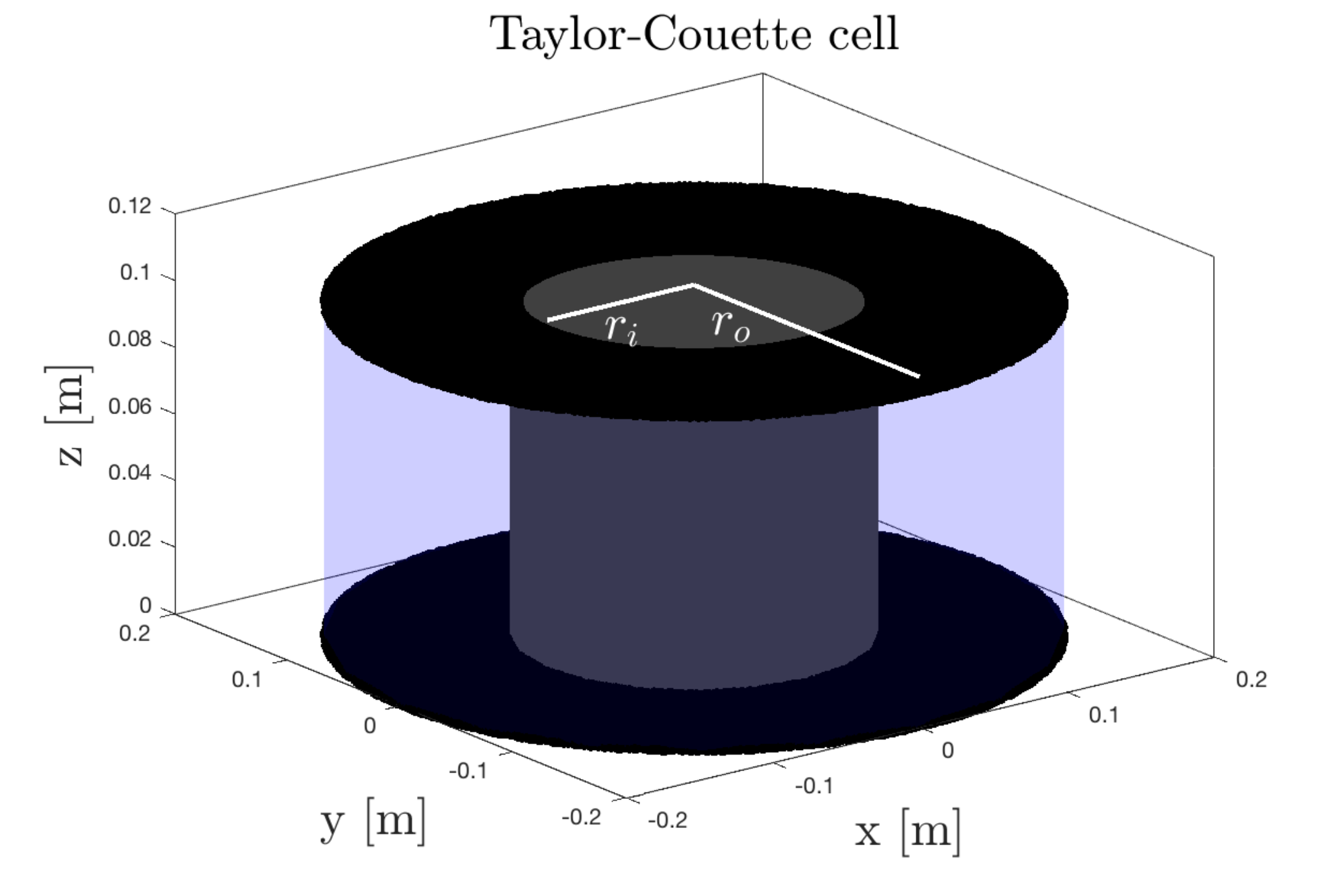}
        \includegraphics[width=.6\textwidth]{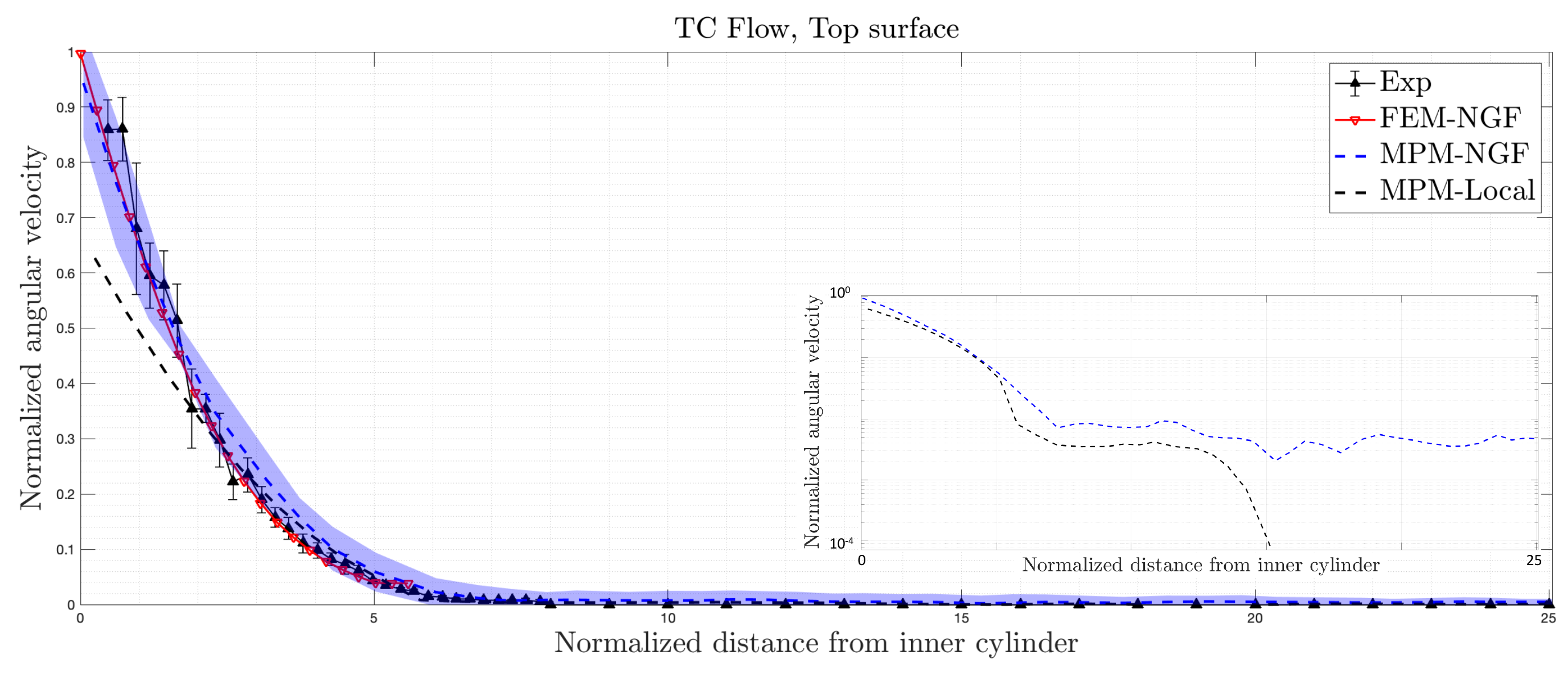}
        \caption{Left: Taylor Couette cell, with chosen top reference surface. The inner cylinder is moving and the outer cylinder is fixed. Right: Normalized angular velocity versus normalized distance from inner cylinder from experiment, FEM-NGF, MPM-NGF, and MPM-Local at top surface in Earth gravity. Inset: Log-scale MPM-NGF and MPM-Local velocity profiles.} 
        % Zoomed-in MPM velocity profile after 5 seconds.}
        \label{fig:tc}
    \end{figure*}   
    
    Here, we emphasize that Excavation is an appropriate case for verifying our hyperelastic MPM implementation of the dynamical nonlocal model (i.e. NGF). While the necessity of nonlocality will be discussed in \S \ref{Excavation} (and \S \ref{TC}), in terms of hyperelasticity, figure \ref{fig:hyper1} shows the distribution of the elastic strain evolution in Excavation (and TC Flow). Note that the strains of the free and rigid particles have been excluded from the means, and the colored areas represent the upper one standard deviation from the means. The large elastic strain in Excavation makes this case appropriate for verifying our hyperelastic implementation. In this case, in fact, the confining pressure is increasing in time as the soil is being accumulated in front of the excavator (i.e. blade). Although the \textit{steady} mean elastic strains in TC Flow are not as large, we will use this test case to address other aspects of our MPM-NGF implementation (similar to \cite{Fredini2013EvaluationOW}, and \cite{hypFemK} which uses TC Flow to verify a hyperelastic FEM implementation of NGF). It will be shown in \S \ref{TC} that the TC Flow solution converges to a correct steady state solution. The elastic strains for the Wheel case stayed low throughout the simulations.
    
    Furthermore, we verify our implementation by tracking the $\text{det}(F^p)$ versus time, and by showing spatial discretization error convergence for the steady TC Flow and unsteady Excavation. Figure \ref{fig:hyper2} shows that the plastic deformation gradient remains within bounds in both TC Flow and Excavation (even as the elastic deformation is increasing in Excavation as shown in figure \ref{fig:hyper1}). Figure \ref{fig:meshconv} indicates the convergence of the simulations toward the experimental results as the normalized spatial discretization size, $\Delta x / (d*\text{ppc})$, is decreased. The convergence errors are measured with respect to our measured force data for the Excavation case, and with respect to published velocity profile data \cite{exp1} for the TC Flow case. The grid size, $\Delta x$, is normalized by grain diameter, $d$ (kept constant within each test case, 0.3 mm for Excavation and 4 mm for TC Flow). Further, particle per cell (ppc; nondimensional) is also included in the denominator because increasing ppc is another way to achieve a finer resolution. For the TC Flow case, ppc is kept constant at 4; for the Excavation case, ppc is set to either 8 or 1, as varying $\Delta x$ and varying ppc were both explored in our prior work \cite{Hae20isarc}.
    
    The NGF parameters shown in table \ref{tab:ngfpar} are set based on the physical properties in our experiments and previous work \cite{DunMScPaper} except when noted otherwise. Also, the MPM spatial and temporal discretization parameters (ppc, $\Delta x$ and $\Delta t$) are selected based on the study in \cite{mpmHu}. However, here some of the parameters are subject to the Von Neumann stability condition of the NGF model's PDE (Poisson equation) where $\Delta t < (\Delta x^2 t_0)/\allowbreak(2A^2d^2)$ \cite{dunPhdK}. Moreover, the computational time spent in all the MPM-NGF simulation models are provided in table \ref{tab:mpmngfspec}. MPM runs can be substantially sped up by reducing the number of particles per cell (ppc) and/or increasing the grid size ($\Delta x$), which all result in a lower particle count. Using a relaxed Young's modulus and a larger time-step ($\Delta t$) would also contribute to this while satisfying the stability conditions of elasticity and the NGF model's PDE \cite{DunMScPaper,DunDemExpK,hypFemK}. An implicit MPM solver could rectify this and is a possible direction of future work. The mean percentage error (MPE) of the excavation forward force, wheel load, and TC flow angular velocity are also specified relative to the experiments.
        
    The simulations were performed on an Intel Core i7-6700 3.4 GHz Quad-Core processor. Houdini Apprentice \cite{houdini} is utilized for the visualizations. From a high performance computing (HPC) point of view, multi-threading (via Intel TBB) and vectorization (via explicit SIMD) were utilized in the code, in addition to some algorithmic improvements explained in \S \ref{subsec:mpm}. These techniques make the current MPM 2x faster than a traditional MPM \cite{mpmHu}. Although the NGF model appends an extra PDE to the governing equations, our numerical approach does not significantly increase the simulation run-time ($\approx5\%$, relative to algebraic or parametric models).

    % ----------
    % (for more details see \cite{Hae20eas})
    % TC Flow    
    \subsection{Taylor-Couette Flow} \label{TC}
\begin{figure}
        \centering
        \includegraphics[width=.48\textwidth]{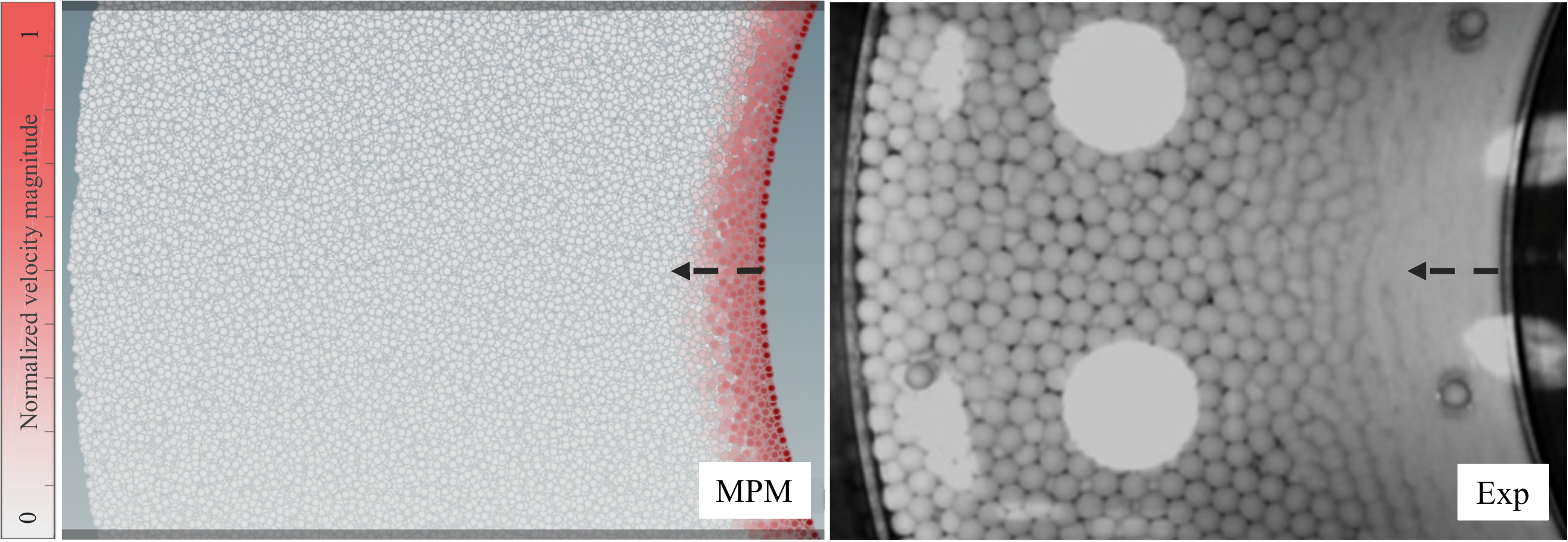}
        \caption{Shear band illustration  based on particle normalized velocity magnitude at the 22 second mark from MPM (left) versus experiment (right) \cite{exp1} at top surface in Earth gravity.}
        \label{fig:1mgshearb}
    \end{figure}

        We primarily verify our approach (i.e. MPM-NGF) using Taylor-Couette Flow via 1) the experimental results that have been produced by a TC flow apparatus (shown in figure \ref{fig:tc}-left) \cite{exp1,exp2,exp3}, and 2) the finite element method (FEM) with NGF already verified in \cite{hypFemK}. The simulation conditions are from the experiment in the literature; data from the top surface, in Earth gravity, are selected as the relevant validation case as they can also be simulated using FEM-NGF. Figure \ref{fig:tc}-right compares the normalized angular velocity profiles between the experiment, FEM-NGF, MPM-NGF, and MPM-Local. It is apparent that the model can capture absolute values, shear band lengths and relative curves. The inset figure also shows the log-scale MPM mean velocity profile which never goes to 0 due to nonlocality - as opposed to a local model (i.e. NGF with $A=0$). The MPE of the velocity profile in the simulation compared to the experimental result, shown in table \ref{tab:mpmngfspec}, does not exceed 1\% magnitude. There is also a very good agreement between the MPM results at the top surface and the validated FEM results. Moreover, particle motions at the top surface at the 22 second mark (i.e. near the end of the experiment) are shown in figure \ref{fig:1mgshearb}. Colors from white to red represent normalized velocity magnitude. The right one is a long exposure image from the experiment \cite{exp1} with blurred areas indicating moving grains. Both the experiment and MPM clearly show the shear band with an exponential decay at the top surface.
        % \deleted{In both MPM and the experiment, the shear band at top is larger than at the bottom in One-G; while FEM is unable to capture such a discrepancy due to gravity. This case study will be presented in greater detail in a separate forthcoming publication.}

    % ----------
    % Excavation
    \subsection{Excavation} \label{Excavation}
\begin{figure}[!t]
            \centering
            \includegraphics[width=0.48\textwidth]{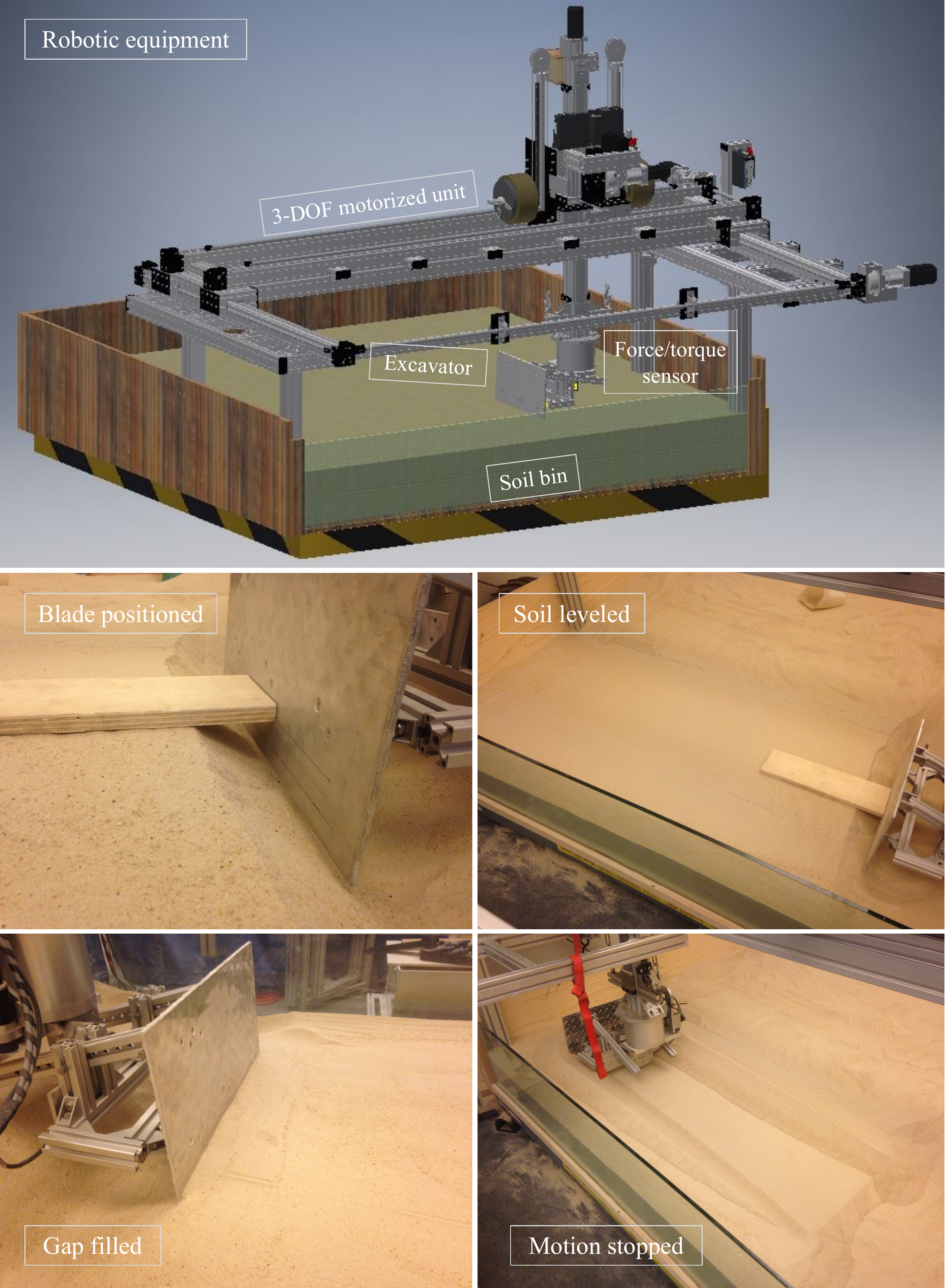}
            \caption{Robotic equipment and experiment setup.}
            \label{fig:excavexp}
        \end{figure}

        We set up an Excavation experiment to validate the MPM with NGF model. It consists of a sandbox positioned under a 3-degree-of-freedom motorized unit to which an excavation tool is attached. For this experiment and for the numerical simulation presented here, the excavation tool is a flat plate (blade) as depicted in figure \ref{fig:excavexp}. The rake angle of the blade can be set manually and it remains constant during the run. The excavator can be moved horizontally and vertically independently. The motors are controlled such that the impulses from the soil flow do not affect the trajectory of the excavator. The excavator is installed on a Delta IP60 (ATI Industrial Automation Inc.) 6-axis force-torque sensor. The blade trajectory is composed of three segments: first, a downward ramped motion at the start to dive into the soil to a specific depth, then a long-duration horizontal motion, and finally an upward ramp at the very end. Two tests were done based on this trajectory but at different (2-cm and 5-cm) depths.
    
        The soil in the experiment is a NASA Glenn Research Center lunar soil simulant GRC-1. The relative density used is 44.6 +/- 7.2\%. This is calculated based on the cone index gradient of 5.30 +/- 0.6 kPa/mm using the correlation in \cite{grc1}. Thus, the corresponding internal friction angle can be obtained as 35 deg. The median grain diameter and density are 0.3 mm and 2583 kg/m\textsuperscript{3}. Using the triaxial test performed by \cite{grc1} the estimated Young's and shear moduli are 15 MPa and 5.8 MPa, respectively. Also, the measured external friction angle between the blade and soil is $\sim$30 deg. The setup of the experiment is shown in figure \ref{fig:excavexp}-left.

        \begin{figure*}[!t]
            \centering
            \includegraphics[width=1\textwidth]{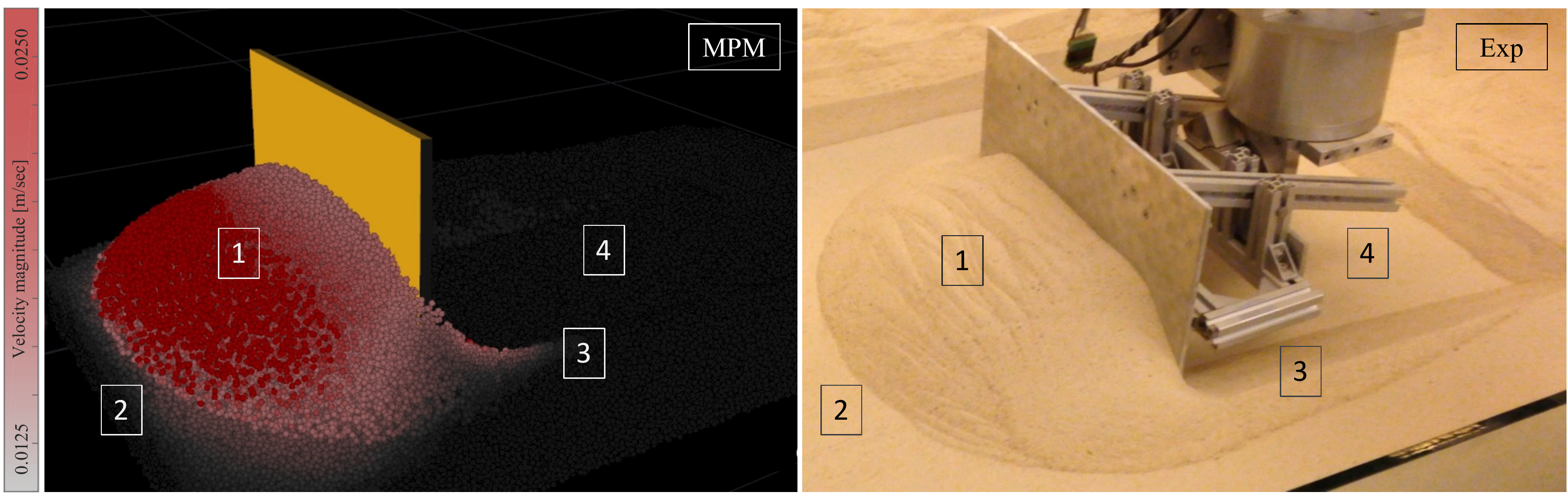}
            \caption{Geometries of soil deformations in MPM-NGF and experiment for 5-cm depth.}
            \label{fig:excavexpmpm}
        \end{figure*}
        
        \begin{figure}
            \centering
            \includegraphics[width=0.48\textwidth]{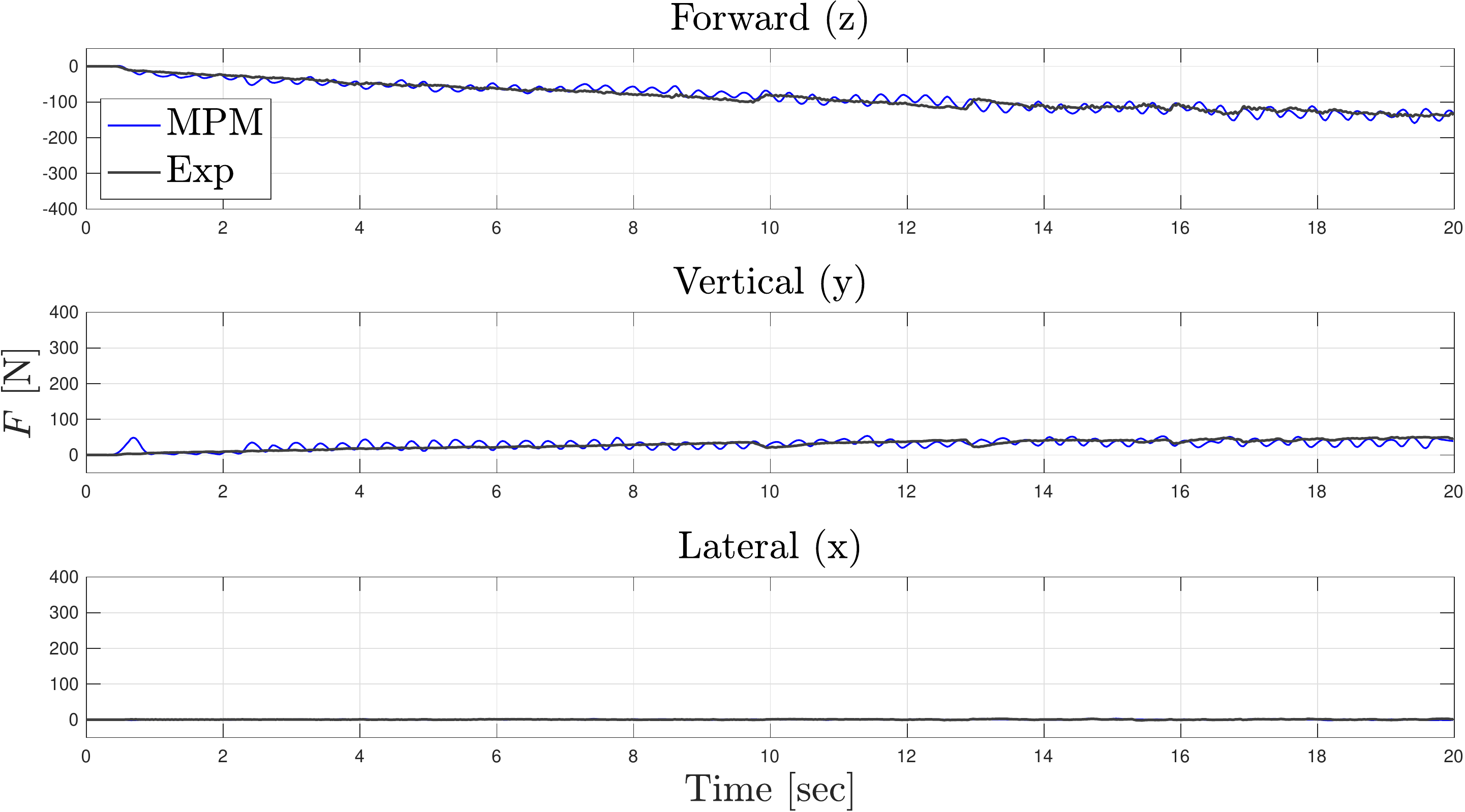}
            \includegraphics[width=0.48\textwidth]{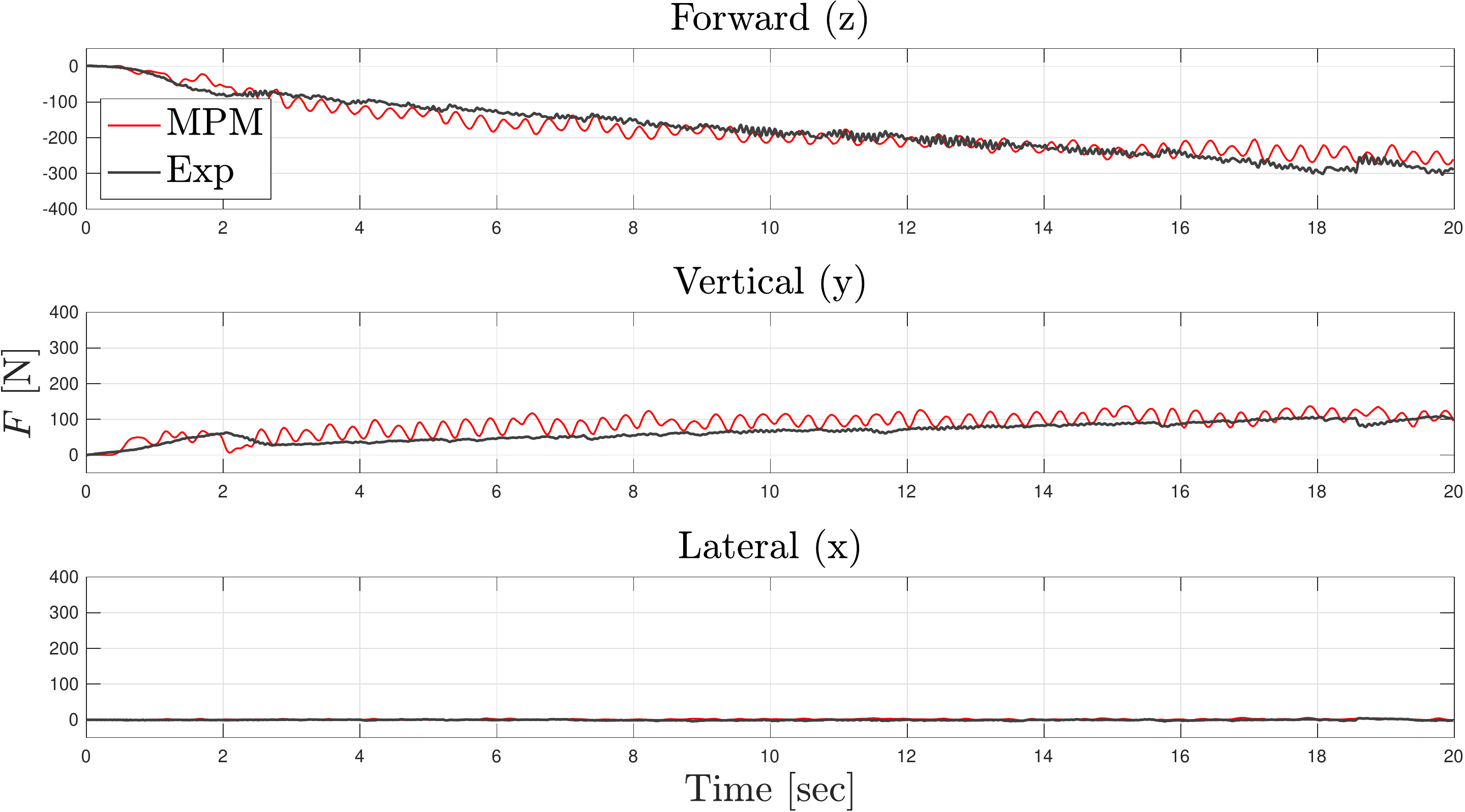}
            \caption{Interaction forces from MPM-NGF and experiment for 2-cm (top) and 5-cm (bottom) depths.}
            \label{fig:force}
        \end{figure}
        
        \begin{figure*}
            \centering
            \includegraphics[width=1\textwidth]{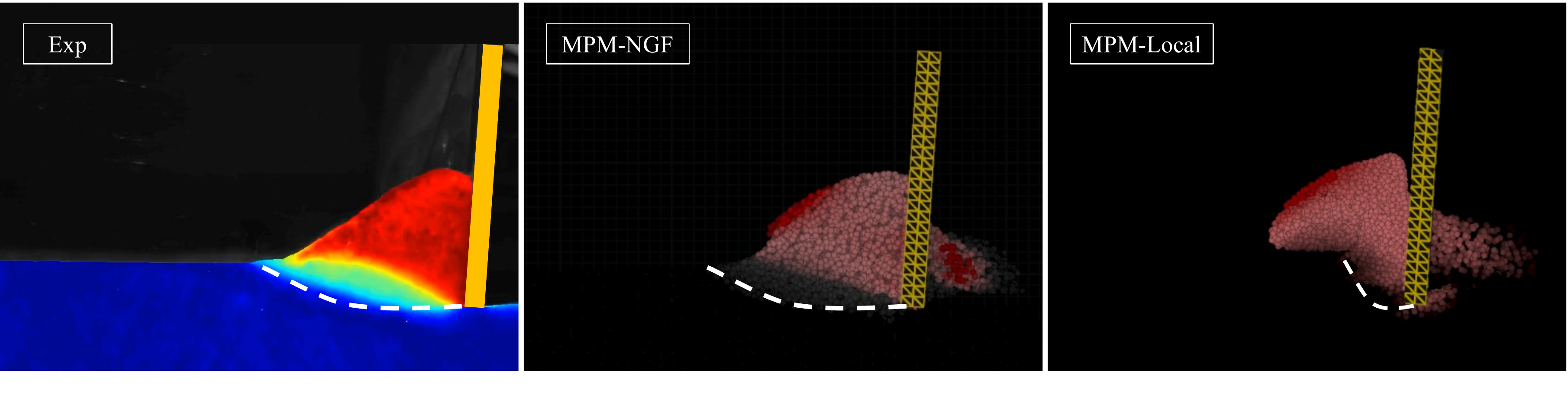}
            \caption{Failure surface in (processed) experiment, MPM with nonlocal model (NGF), and MPM with local model (i.e. $A=0$ in NGF).}
            \label{fig:failureplane}
        \end{figure*}
        
        \begin{figure*}[!t]
            \centering
            \includegraphics[width=1\textwidth]{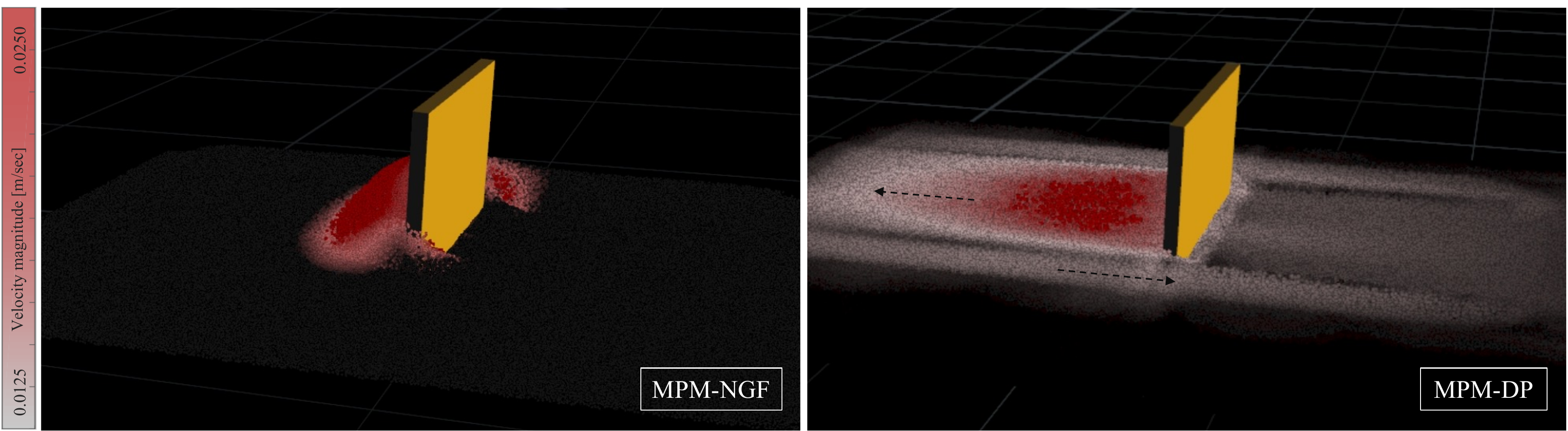}
            \caption{Excavation modeling with MPM-NGF (left) and MPM-DP (right). Soil modeled with DP experiences unexpected fluid-like behaviors in untouched areas with no accumulating pile.}
            \label{fig:excavmpmngfdp}
        \end{figure*}
        
        A qualitative and visual comparison is shown in figure \ref{fig:excavexpmpm}. This shows the soil geometry in the end of the second trajectory segment in the 5-cm MPM-NGF and experiment. Four visual criteria are emphasized for a better analogy. In general, the soil behavior in MPM-NGF is predicted similar to the one in the experiment. The particle velocities visualized in colors clearly depict the static (gray) and dynamic (red) parts. The MPM-NGF velocity field corresponds well with the experiment. Furthermore, the tool-soil interactions in the simulations here are evaluated by the forces measured in the experiments. Figure \ref{fig:force} compares all the forces from MPM-NGF and experiment. The quantitative force values are in good agreement with the experimental forces in the three (forward, vertical and lateral) directions. The MPE of the forward force of the 5-cm depth case in the simulation compared to the experimental result is provided in table \ref{tab:mpmngfspec}. The errors do not exceed 1\% and 16\% magnitude for the original and faster simulations, respectively.
        % Due to the fast technique used to calculate the NGF Laplacian term in MPM,
        Due to the dimensional analysis used to model the large experimental cases, the scaled MPM-NGF results seem to have slightly boosted oscillation amplitude. However, in addition to the overall trend, MPM-NGF is able to capture drops and rises in force at various time-steps of the two experiments. This can highlight the unsteady form of the MPM solver as well as the NGF constitutive model.
        
        The nonlocality effect in MPM is further examined by an experiment with the blade positioned against the glass sidewall. For this experiment, in addition to force data, high-speed images of the  blade-soil interaction were also captured. Soil motion was analyzed from these images using the Soil Optical Flow Technique (SOFT) \cite{kvis}. The  soil flow visualization in Figure \ref{fig:excavmpmngfdp} reveals the soil failure surface \cite{Hae20isarc}. The failure surface has also been captured by MPM with the nonlocal model (NGF). However, the local model (i.e. NGF with $A=0$) does not correctly capture the failure surface. Also, we tried modeling the excavation case with a recent modified Drucker-Prager (DP) plasticity model (with projection) \cite{klarDP}. This model has shown its capability in modeling the Taylor-Couette flow \cite{Hae20eas}. In addition to the material parameters mentioned before, the hardening parameters were set as $h_0=44$, $h_1=9$, $h_2=0.2$, and $h_3=9$. Parameters $h_0$ and $h_3$ are responsible for the minimum and maximum friction angle relative to the plastic deformation (here 35 and 44 degrees, respectively). Also, $h_1$ and $h_2$ define the change rate of the friction angle \cite{refDP}. We found the model insensitive to the depths we were interested in (i.e. 2 cm and 5 cm). The fluid-like behavior of the soil modeled with this plastic model is shown in figure \ref{fig:excavmpmngfdp}-right compared to the soil deformation modeled with the viscoplastic NGF model (left). It seems that the DP model has issue with the hardening part, because as soon as we increased the depth, the soil started to pile up in front of the excavator yet with unexpected fluid-like behaviors in very distant areas, and to the sides of the blade even reflected soil flow in the opposite direction. With the help of visualization of the particle velocities from black (0 m/s) to white (0.0125 m/s) and red (0.0250 m/s), the deformations are clearly visible in figure \ref{fig:excavmpmngfdp}. MPM with DP plasticity would clearly require further development in future work, if there is a chance for it to properly capture the experimental results.

    % ----------
    % Additional: Wheel, Industrial Excavator and Silo
    \subsection{Additional Configurations} \label{Additional}
        % Wheel
        \begin{figure}[!]
            \centering
            \includegraphics[width=0.48\textwidth]{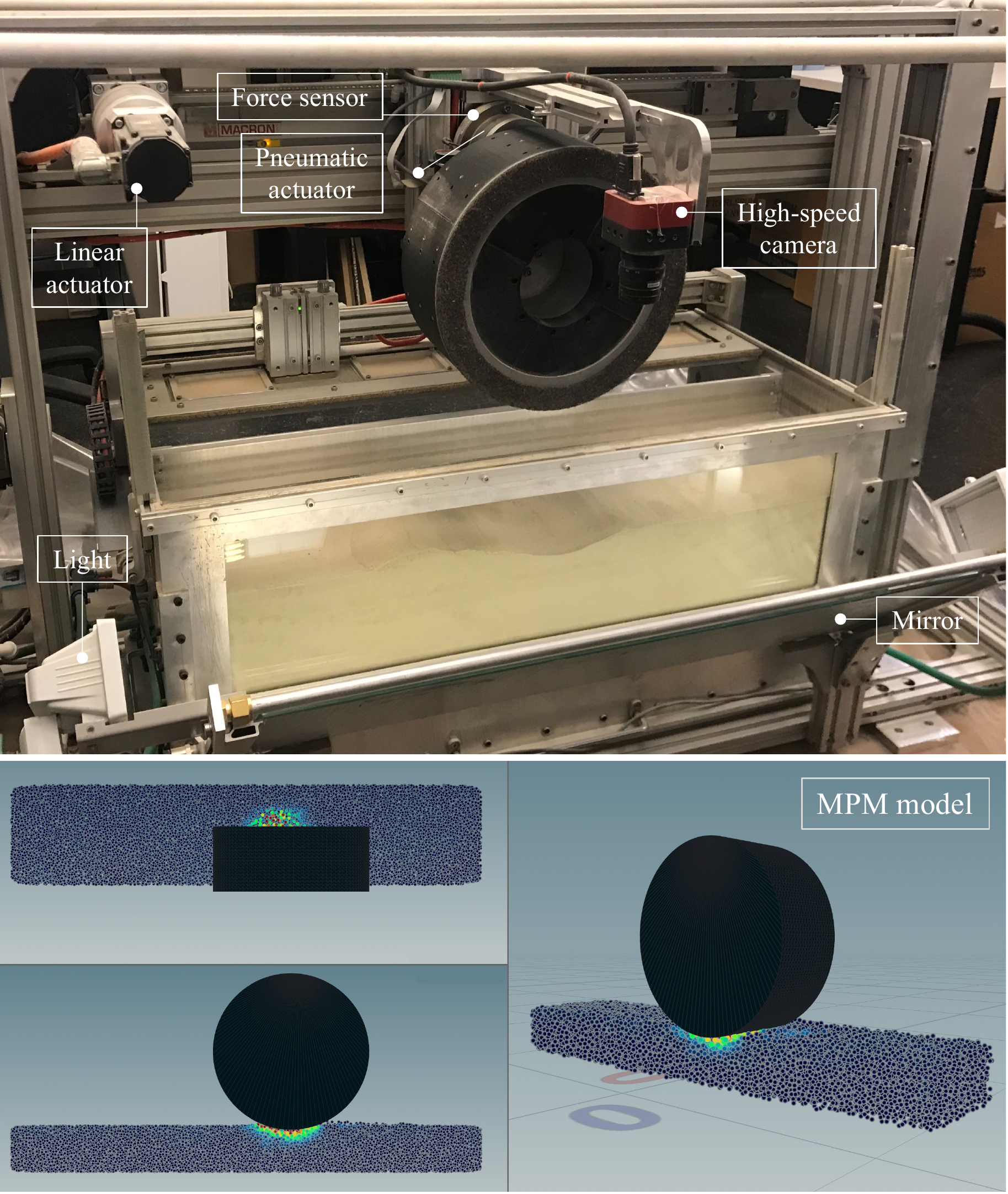}
            \caption{Left: Robotic equipment and experiment setup. Right: Wheel-soil model in MPM.}
            \label{fig:model}
        \end{figure}
        
        \begin{figure*}[!t]
            \centering
            \includegraphics[width=1\textwidth]{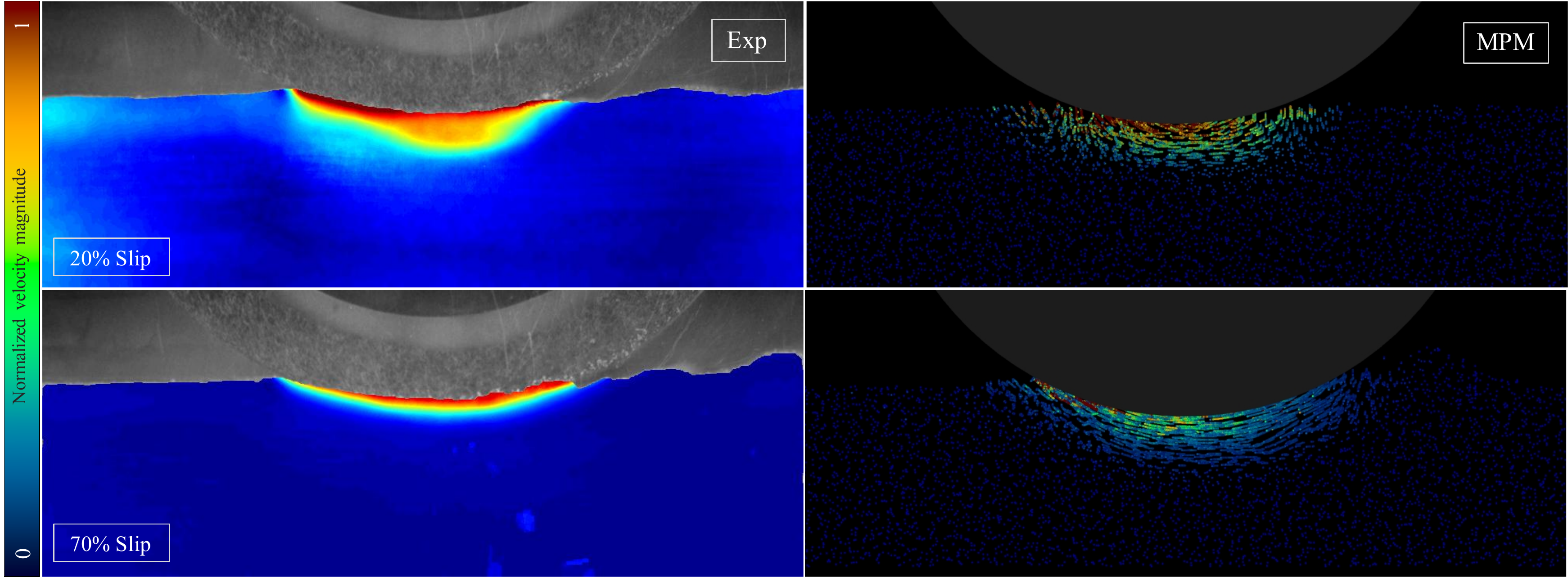}
            \caption{Mean normalized subsurface soil velocity magnitude between 5 and 6 second marks from experiment (left) and MPM-NGF (right) with 20\% (top) and 70\% (bottom) slips.}
            \label{fig:wheelcolor}
        \end{figure*}
        
        \begin{figure*}[!t]
            \centering
            \includegraphics[width=1\textwidth]{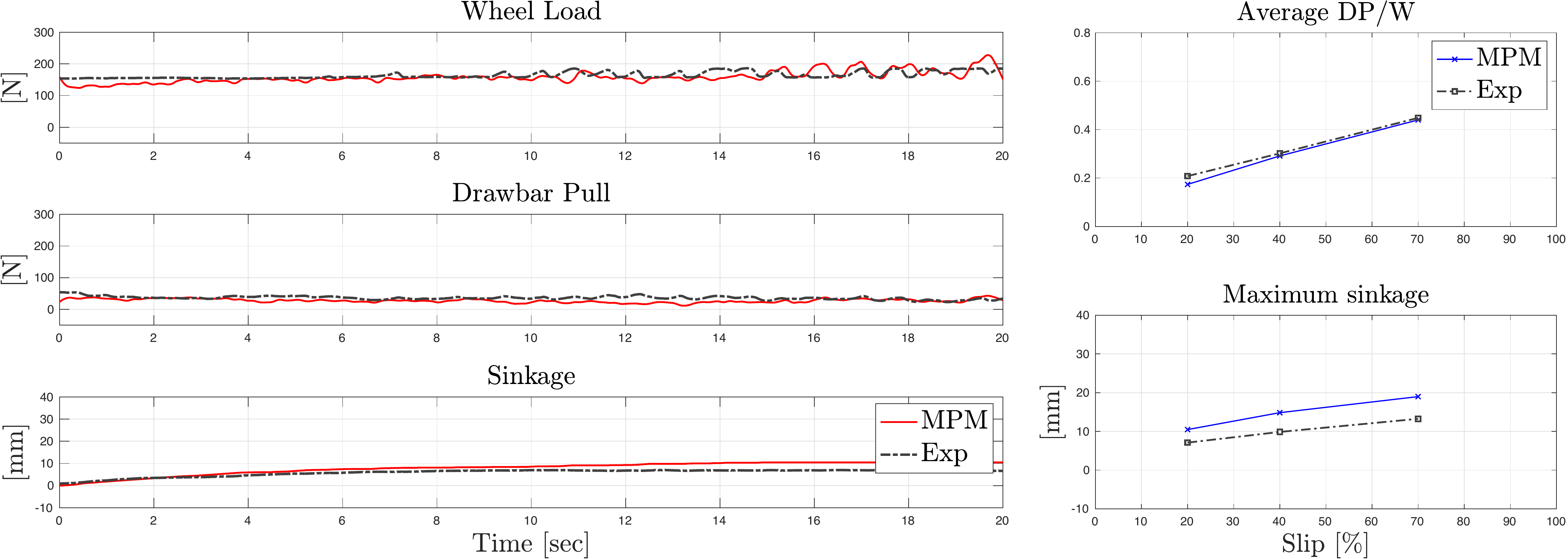}
            \caption{Left: Interaction forces (drawbar pull and wheel load) and sinkage from experiment and MPM-NGF with 20\% slip. Right: Mean drawbar pull to wheel load ratio and maximum sinkage at different slips from experiment and MPM-NGF.}
            \label{fig:res}
        \end{figure*}
        
        \textbf{Wheel.}
        In addition to the previous test cases, we also validate the approach via wheel-on-soil experiments. To this end, our specialized robotic single-wheel apparatus was utilized \cite{Nik20eas}. The rover wheel is driven via synchronized control of a horizontal linear actuator (Macron Dynamics R6S driven by a Kollmorgen AKM23C motor and AKD-P00306 driver) and a wheel motor (Maxon RE35 with a MaxPos 50/5 driver) in an instrumented soil bin with a dimension of 90 x 19.3 x 35 cm. A vertical load is applied to the wheel while allowing free vertical motion. Data collection includes forces in the forward (drawbar pull, DP, i.e. net traction) and vertical (wheel load, W) directions via an ATI Delta IP60 sensor, and vertical wheel displacement via an ALPS slide potentiometer. The wheel is located against a transparent window in the soil bin. A high-speed 4 MP monochrome camera with 16 mm EFL lens and Core2 digital video recorder observes wheel–soil interactions through a glass sidewall in the soil bin via a mirror tilted at 45 degrees. The images are captured, processed, and analyzed using SOFT \cite{kvis}. Experiment setup in the apparatus and its model in MPM is shown in figure \ref{fig:model}.
        
        The smooth wheel is 30 cm in diameter and 12.5 cm in width. The soil in the experiment is the GRC-1 with the aforementioned properties. Also, the measured external friction coefficient between the wheel and soil is approximately 0.4. The amount of traction a wheel produces is related to how much it slips. Thus, we tested with three different slips including $S = 20\%$, 40\%, and 70\% where $S=100(1-v/r\omega)$ is a function of wheel radius $r$, horizontal speed $v$, and rotational speed $\omega$. Here, we varied the horizontal velocity to change the amount of slip. Moreover, we exerted a normal load of 164 N on the wheel. This load consists of the wheel unit weight plus applied force on the wheel unit.
    
        A visual quantitative comparison is shown in figure \ref{fig:wheelcolor}. This shows the mean subsurface soil motion between the 5 and 6 second marks in the experiment and simulation (MPM-NGF) with 20\% and 70\% slips. The grain/particle velocities are normalized with the tangential velocity of the wheel rim. The velocities visualized in colors clearly depict the static (blue) and highly dynamic (red) parts. The soil deformation underneath and behind the wheel (right side) from experiment and simulation are comparable.
    
        \begin{figure*}[!t]
            \centering
            \includegraphics[width=1\textwidth]{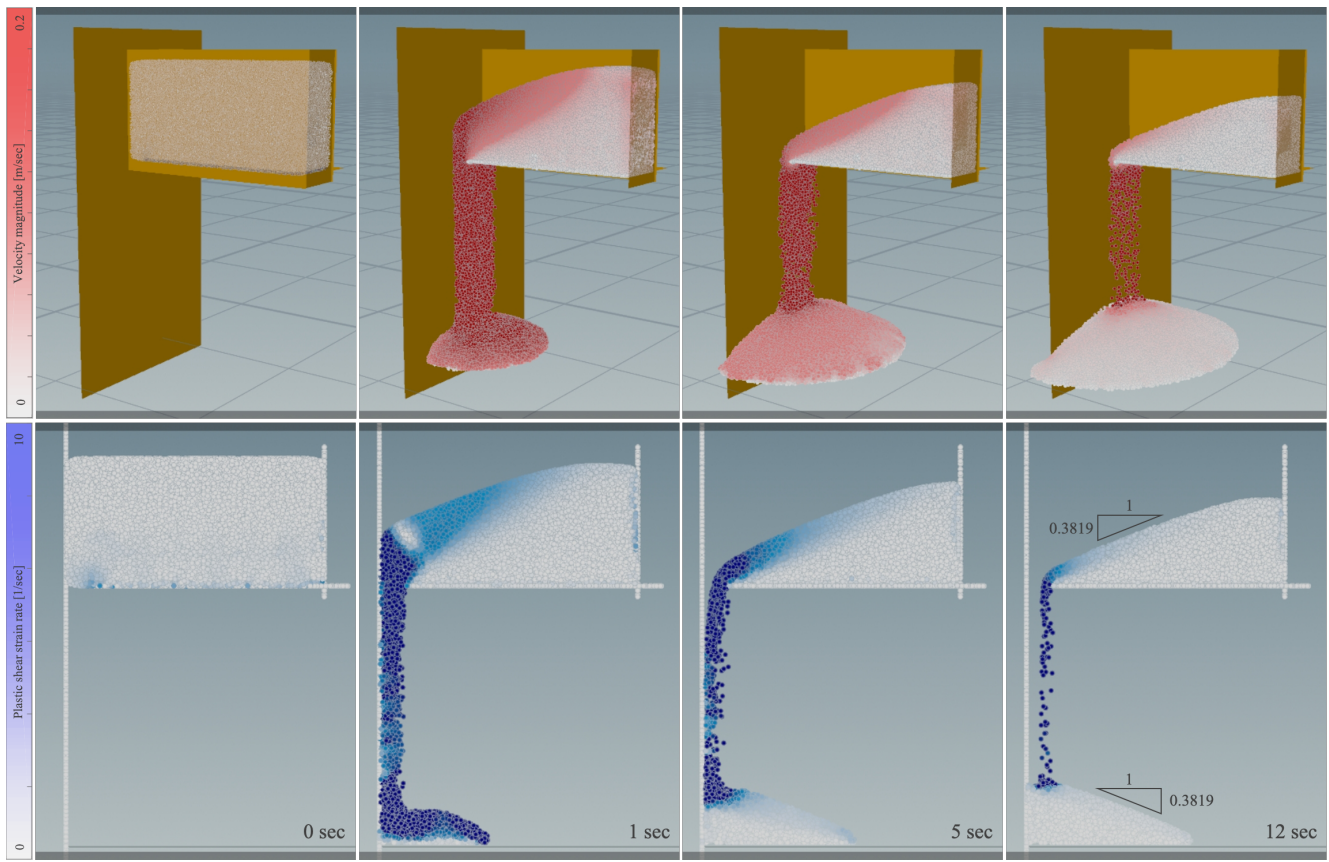}
            \caption{Top: Velocity filed in 3D silo. Bottom: Plastic shear strain rate field shown in a thin layer of 3D silo.}
            \label{fig:silo}
        \end{figure*}
    
        Furthermore, the wheel-soil interactions from simulation here are evaluated by the forces measured in the experiments. Figure \ref{fig:res}-left compares the drawbar pull and wheel load forces from MPM-NGF and experiment with 20\% slip. The quantitative force values are in good agreement with the experimental forces in both the forward and vertical directions. The MPE of the wheel load force in the simulation compared to the experimental result, provided in table \ref{tab:mpmngfspec}, does not exceed 6\% magnitude. In addition to the overall trend here, as mentioned, MPM-NGF is also able to capture drops and rises in force at various time-steps. The NGF parameters are set based on the physical properties in the experiment. In the simulation, to settle the wheel quickly and prevent wheel vertical bouncing, we applied an exponential damping to the vertical motion of the wheel for one second prior to start. The wheel upward motion was also significantly damped out during the wheel forward motion. This means, the wheel could move only downward (but not upward) freely.
    
        Moreover, figure \ref{fig:res}-right compares the mean drawbar pull to wheel load ratio (DP/W) and maximum sinkage at different slips from experiment and MPM-NGF. The DP/W ratio is a common metric in the field of terramechanics that refers to the load a vehicle can tow relative to vertical load. Wheel sinkage, as a function of slip, is also a measure of the response of the terrain to a specific loading, and affects wheel performance. The simulation DP/W ratio is in a very good agreement with the experiment. Also, due to the implementation of the wheel upward damping, the maximum sinkage is slightly higher in the simulation than in the experiment. However, MPM-NGF has been able to capture the overall trends.
    
        % Industrial Excavator
        \textbf{Industrial Excavator.}
        We further model an industrial excavation with MPM-NGF shown in figure \ref{fig:bucket}. It demonstrates the ability of MPM-NGF to model the three states that a granular material can experience during soil cutting operations \cite{Hae20isarc}. The colors indicate the equivalent plastic shear strain rate ($\dot{\gamma}^\text{p}$) from black (0 $\text{s}^{-1}$) to light green (5 $\text{s}^{-1}$) for each particle. This enables us to track the soil deformations in the particle's body frame as opposed to the velocity magnitude in the global frame. At the 6 second mark, the soil inside the bucket and untouched areas are under solid-like elastic deformation. The separated particles leaking from the bucket are under gas-like stress-free deformation. Finally, the particles piled up after returning to the bin are under fluid-like viscoplastic deformation.

        % Silo
        \textbf{3D Silo.}
        Finally, similar to recent work in the literature \cite{DunMScPaper}, we model the drainage of a silo here but via the NGF model and in 3D. The material parameters are given in table \ref{tab:ngfpar} except for $\mu_\text{s}=0.3819$. The bottom boundary is rough and sides are smooth. This case again demonstrates the fact that the model can capture the three states of matter. Figure \ref{fig:silo}(top), a perspective 3D view of the silo, shows the velocity magnitude of each individual particle from white (0 m/s) to red (0.2 m/s) at 0, 1, 5, and 12 second marks. Also, a side 2D view of a thin layer in the middle is shown in figure \ref{fig:silo}(bottom); it visualizes the equivalent plastic shear strain rate from white (0 $\text{s}^{-1}$) to blue (10 $\text{s}^{-1}$). The slope of the soil pile at the 12 second mark matching the friction angle provides further evidence of the accuracy of the model.

\section{Conclusion}
    A numerical approach was designed to implement the dynamical and hyperelastic form of the nonlocal granular fluidity constitutive model in three-dimensional material point method. The approach was validated by our experiments and various configurations in literature including excavation, wheel-soil, silo, and Taylor-Couette flow. Future work can be the implementation of an implicit and coupled approach/scheme for solving the governing PDEs (momentum and nonlocal equations). If successful, this would lead to the numerical approach to being unconditionally stable. Furthermore, appending the recently developed transient granular rheology \cite{Kam19trans} would enable the approach to model the unsteady pre-critical state behaviors in granular flows.

\begin{acknowledgements}
    The authors gratefully acknowledge f-\\unding for this work from the Natural Sciences and Engineering Research Council of Canada (RGPIN-2015-06046).
\end{acknowledgements}


% \bibliographystyle{spmpsci}      % mathematics and physical sciences
\bibliographystyle{spphys}       % APS-like style for physics
\bibliography{refs}   % name your BibTeX data base

\begin{thebibliography}{10}
\providecommand{\url}[1]{{#1}}
\providecommand{\urlprefix}{URL }
\expandafter\ifx\csname urlstyle\endcsname\relax
  \providecommand{\doi}[1]{DOI \discretionary{}{}{}#1}\else
  \providecommand{\doi}{DOI \discretionary{}{}{}\begingroup
  \urlstyle{rm}\Url}\fi

\bibitem{Kam17complex}
{Kamrin, Ken}, EPJ Web Conf. \textbf{140}, 01007 (2017).
\newblock \doi{10.1051/epjconf/201714001007}.
\newblock \urlprefix\url{https://doi.org/10.1051/epjconf/201714001007}

\bibitem{Cun79dem}
P.A. Cundall, O.D.L. Strack, Géotechnique \textbf{29}(1), 47 (1979).
\newblock \doi{10.1680/geot.1979.29.1.47}.
\newblock \urlprefix\url{https://doi.org/10.1680/geot.1979.29.1.47}

\bibitem{comp13}
H.~Zhu, Y.~Wu, A.~Yu, China Particuology \textbf{3}(6), 354 (2005).
\newblock \doi{https://doi.org/10.1016/S1672-2515(07)60215-2}.
\newblock
  \urlprefix\url{https://www.sciencedirect.com/science/article/pii/S1672251507602152}

\bibitem{DunDemExpK}
S.~{Dunatunga}, K.~{Kamrin}, Journal of Mechanics Physics of Solids
  \textbf{100}, 45 (2017).
\newblock \doi{10.1016/j.jmps.2016.12.002}.
\newblock \urlprefix\url{https://ui.adsabs.harvard.edu/abs/2017JMPSo.100...45D}

\bibitem{Mac14flex}
M.~Macklin, M.~M\"{u}ller, N.~Chentanez, T.Y. Kim, ACM Trans. Graph.
  \textbf{33}(4) (2014).
\newblock \doi{10.1145/2601097.2601152}.
\newblock \urlprefix\url{https://doi.org/10.1145/2601097.2601152}

\bibitem{Hol14p2}
D.~Holz, in \emph{Workshop on Virtual Reality Interaction and Physical
  Simulation}, ed. by J.~Bender, C.~Duriez, F.~Jaillet, G.~Zachmann (The
  Eurographics Association, 2014).
\newblock \doi{10.2312/vriphys.20141232}

\bibitem{Bou14pd}
S.~Bouaziz, S.~Martin, T.~Liu, L.~Kavan, M.~Pauly, ACM Trans. Graph.
  \textbf{33}(4) (2014).
\newblock \doi{10.1145/2601097.2601116}.
\newblock \urlprefix\url{https://doi.org/10.1145/2601097.2601116}

\bibitem{comp14n1}
R.~Irani, R.~Bauer, A.~Warkentin, Journal of Terramechanics \textbf{48}(4), 307
  (2011).
\newblock \doi{https://doi.org/10.1016/j.jterra.2011.05.001}.
\newblock
  \urlprefix\url{https://www.sciencedirect.com/science/article/pii/S0022489811000334}

\bibitem{comp15new}
C.~Senatore, K.~Iagnemma, Journal of Terramechanics \textbf{51}, 1 (2014).
\newblock \doi{https://doi.org/10.1016/j.jterra.2013.10.003}.
\newblock
  \urlprefix\url{https://www.sciencedirect.com/science/article/pii/S0022489813000797}

\bibitem{comp14n2}
L.~Ding, Z.~Deng, H.~Gao, J.~Tao, K.D. Iagnemma, G.~Liu, Journal of Field
  Robotics \textbf{32}(6), 827 (2015).
\newblock \doi{10.1002/rob.21533}.
\newblock \urlprefix\url{https://doi/abs/10.1002/rob.21533}

\bibitem{God14cm}
J.D. Goddard, Applied Mechanics Reviews \textbf{66}(5) (2014).
\newblock \doi{10.1115/1.4026242}.
\newblock \urlprefix\url{https://doi.org/10.1115/1.4026242}.
\newblock 050801

\bibitem{am}
P.Y. Lagrée, D.~Lhuillier, European Journal of Mechanics - B/Fluids
  \textbf{25}(6), 960 (2006).
\newblock \doi{https://doi.org/10.1016/j.euromechflu.2006.03.003}.
\newblock
  \urlprefix\url{https://www.sciencedirect.com/science/article/pii/S0997754606000331}

\bibitem{DunMScPaper}
S.~Dunatunga, K.~Kamrin, Journal of Fluid Mechanics \textbf{779}, 483–513
  (2015).
\newblock \doi{10.1017/jfm.2015.383}

\bibitem{comp12}
X.~Zhang, Z.~Chen, Y.~Liu, \emph{The material point method} (Elsevier, 2014).
\newblock
  \urlprefix\url{https://www.elsevier.com/books/the-material-point-method/zhang/978-0-12-407716-4}

\bibitem{mpmHu}
Y.~Hu, Y.~Fang, Z.~Ge, Z.~Qu, Y.~Zhu, A.~Pradhana, C.~Jiang, ACM Trans. Graph.
  \textbf{37}(4) (2018).
\newblock \doi{10.1145/3197517.3201293}.
\newblock \urlprefix\url{https://doi.org/10.1145/3197517.3201293}

\bibitem{klarDP}
G.~Kl\'{a}r, T.~Gast, A.~Pradhana, C.~Fu, C.~Schroeder, C.~Jiang, J.~Teran, ACM
  Trans. Graph. \textbf{35}(4) (2016).
\newblock \doi{10.1145/2897824.2925906}.
\newblock \urlprefix\url{https://doi.org/10.1145/2897824.2925906}

\bibitem{Hae20isarc}
A.~Haeri, D.~Tremblay, K.~Skonieczny, D.~Holz, M.~Teichmann, in
  \emph{Proceedings of the 37th International Symposium on Automation and
  Robotics in Construction (ISARC)} (International Association for Automation
  and Robotics in Construction (IAARC), Kitakyushu, Japan, 2020), pp. 608--615.
\newblock \doi{10.22260/ISARC2020/0085}

\bibitem{Hae20eas}
A.~Haeri, K.~Skonieczny, \emph{Granular Flow Modeling of Robot-Terrain
  Interactions in Reduced Gravity}, pp. 51--61.
\newblock \doi{10.1061/9780784483374.006}.
\newblock
  \urlprefix\url{https://ascelibrary.org/doi/abs/10.1061/9780784483374.006}

\bibitem{Mon92sph}
J.J. Monaghan, Annual Review of Astronomy and Astrophysics \textbf{30}(1), 543
  (1992).
\newblock \doi{10.1146/annurev.aa.30.090192.002551}.
\newblock \urlprefix\url{https://doi.org/10.1146/annurev.aa.30.090192.002551}

\bibitem{Bri15flip}
R.~Bridson, CRC Press  (2015).
\newblock \doi{10.1201/b10635}

\bibitem{Har64pic}
F.H. Harlow, Methods Comput. Phys. \textbf{3}, 319–343 (1964)

\bibitem{apic}
C.~Jiang, C.~Schroeder, A.~Selle, J.~Teran, A.~Stomakhin, ACM Trans. Graph.
  \textbf{34}(4) (2015).
\newblock \doi{10.1145/2766996}.
\newblock \urlprefix\url{https://doi.org/10.1145/2766996}

\bibitem{polypic}
C.~Fu, Q.~Guo, T.~Gast, C.~Jiang, J.~Teran, ACM Trans. Graph. \textbf{36}(6)
  (2017).
\newblock \doi{10.1145/3130800.3130878}.
\newblock \urlprefix\url{https://doi.org/10.1145/3130800.3130878}

\bibitem{Kam10regime}
K.~Kamrin, International Journal of Plasticity \textbf{26}(2), 167 (2010).
\newblock \doi{https://doi.org/10.1016/j.ijplas.2009.06.007}.
\newblock
  \urlprefix\url{https://www.sciencedirect.com/science/article/pii/S0749641909000898}

\bibitem{comp16}
A.~Schoefield, P.~Wroth, Pergamon, Oxford  (1968)

\bibitem{comp17}
R.M. Nedderman, \emph{Statics and Kinematics of Granular Materials} (Cambridge
  University Press, 1992).
\newblock \doi{10.1017/CBO9780511600043}

\bibitem{comp20}
S.A. Elaskar, L.A. Godoy, International Journal of Mechanical Sciences
  \textbf{40}(10), 1001 (1998).
\newblock \doi{https://doi.org/10.1016/S0020-7403(98)00004-6}.
\newblock
  \urlprefix\url{https://www.sciencedirect.com/science/article/pii/S0020740398000046}

\bibitem{comp15}
C.K.K. Lun, S.B. Savage, D.J. Jeffrey, N.~Chepurniy, Journal of Fluid Mechanics
  \textbf{140}, 223–256 (1984).
\newblock \doi{10.1017/S0022112084000586}

\bibitem{DP}
D.C. DRUCKER, W.~PRAGER, Quarterly of Applied Mathematics \textbf{10}(2), 157
  (1952).
\newblock \urlprefix\url{http://www.jstor.org/stable/43633942}

\bibitem{naflowRule}
J.~Bonet, A.J. Gil, R.D. Wood, \emph{Nonlinear Solid Mechanics for Finite
  Element Analysis: Statics} (Cambridge University Press, 2016).
\newblock \doi{10.1017/CBO9781316336144}

\bibitem{local}
P.~Jop, Y.~Forterre, O.~Pouliquen, Letters \textbf{441} (2006).
\newblock \doi{10.1038/nature04801}

\bibitem{3dStNonK}
D.L. Henann, K.~Kamrin, Proceedings of the National Academy of Sciences
  \textbf{110}(17), 6730 (2013).
\newblock \doi{10.1073/pnas.1219153110}.
\newblock \urlprefix\url{https://www.pnas.org/content/110/17/6730}

\bibitem{soilp}
W.F. Chen, G.Y. Baladi, in \emph{Soil plasticity: Theory and implementation}
  (Elsevier, 1985)

\bibitem{mpmCourse}
C.~Jiang, C.~Schroeder, J.~Teran, A.~Stomakhin, A.~Selle, in \emph{ACM SIGGRAPH
  2016 Courses} (Association for Computing Machinery, New York, NY, USA, 2016),
  SIGGRAPH '16.
\newblock \doi{10.1145/2897826.2927348}.
\newblock \urlprefix\url{https://doi.org/10.1145/2897826.2927348}

\bibitem{dunMScK}
S.~Dunatunga, A nonlocal dense granular flow model implemented in the material
  point method.
\newblock {MSc} dissertation, Massachusetts Institute of Technology (2014)

\bibitem{soilp2}
W.S. Abdullah, Journal of Civil Engineering \textbf{5} (2011)

\bibitem{nonlocal}
K.~Kamrin, Frontiers in Physics \textbf{7}, 116 (2019).
\newblock \doi{10.3389/fphy.2019.00116}.
\newblock
  \urlprefix\url{https://www.frontiersin.org/article/10.3389/fphy.2019.00116}

\bibitem{Kam18size}
K.~Kamrin, \emph{Quantitative Rheological Model for Granular Materials: The
  Importance of Particle Size} (Springer International Publishing, Cham, 2020),
  pp. 153--176.
\newblock \doi{10.1007/978-3-319-44680-6_148}.
\newblock \urlprefix\url{https://doi.org/10.1007/978-3-319-44680-6_148}

\bibitem{2dStNonK}
K.~Kamrin, G.~Koval, Phys. Rev. Lett. \textbf{108}, 178301 (2012).
\newblock \doi{10.1103/PhysRevLett.108.178301}.
\newblock
  \urlprefix\url{https://link.aps.org/doi/10.1103/PhysRevLett.108.178301}

\bibitem{UnsNonK}
D.L. Henann, K.~Kamrin, International Journal of Plasticity \textbf{60}, 145
  (2014).
\newblock \doi{https://doi.org/10.1016/j.ijplas.2014.05.002}.
\newblock
  \urlprefix\url{https://www.sciencedirect.com/science/article/pii/S0749641914000989}

\bibitem{Ara01pft}
I.S. Aranson, L.S. Tsimring, Phys. Rev. E \textbf{64}, 020301 (2001).
\newblock \doi{10.1103/PhysRevE.64.020301}.
\newblock \urlprefix\url{https://link.aps.org/doi/10.1103/PhysRevE.64.020301}

\bibitem{Ara02pft}
I.~Aranson, L.~Tsimring, Phys Rev E.  (2002).
\newblock \doi{10.1103/PhysRevE.65.061303}

\bibitem{Vol03pft}
D.~Volfson, L.S. Tsimring, I.S. Aranson, Phys. Rev. Lett. \textbf{90}, 254301
  (2003).
\newblock \doi{10.1103/PhysRevLett.90.254301}.
\newblock
  \urlprefix\url{https://link.aps.org/doi/10.1103/PhysRevLett.90.254301}

\bibitem{Bou13igm}
M.~Bouzid, M.~Trulsson, P.~Claudin, E.~Cl\'ement, B.~Andreotti, Phys. Rev.
  Lett. \textbf{111}, 238301 (2013).
\newblock \doi{10.1103/PhysRevLett.111.238301}.
\newblock
  \urlprefix\url{https://link.aps.org/doi/10.1103/PhysRevLett.111.238301}

\bibitem{Bou15igm}
M.~Bouzid, M.~Trulsson, E.~Clément, P.~Claudin, B.~Andreotti, Eur Phys J E.
  \textbf{38}, 1–15 (2015).
\newblock \doi{10.1140/epje/i2015-15125-1}

\bibitem{Pou09sam}
O.~Pouliquen, Y.~Forterre, Philosophical Transactions of the Royal Society A:
  Mathematical, Physical and Engineering Sciences \textbf{367}(1909), 5091
  (2009).
\newblock \doi{10.1098/rsta.2009.0171}.
\newblock
  \urlprefix\url{https://royalsocietypublishing.org/doi/abs/10.1098/rsta.2009.0171}

\bibitem{hypFemK}
D.L. Henann, K.~Kamrin, International Journal for Numerical Methods in
  Engineering \textbf{108}(4), 273 (2016).
\newblock \doi{https://doi.org/10.1002/nme.5213}.
\newblock
  \urlprefix\url{https://onlinelibrary.wiley.com/doi/abs/10.1002/nme.5213}

\bibitem{dunPhdK}
S.~Dunatunga, A framework for continuum simulation of granular flow.
\newblock {PhD} dissertation, Massachusetts Institute of Technology (2017)

\bibitem{cteVF}
P.~JOP, Y.~FORTERRE, O.~POULIQUEN, Journal of Fluid Mechanics \textbf{541},
  167–192 (2005).
\newblock \doi{10.1017/S0022112005005987}

\bibitem{spgrid}
R.~Setaluri, M.~Aanjaneya, S.~Bauer, E.~Sifakis, ACM Trans. Graph.
  \textbf{33}(6) (2014).
\newblock \doi{10.1145/2661229.2661269}.
\newblock \urlprefix\url{https://doi.org/10.1145/2661229.2661269}

\bibitem{pdsampler}
R.~Bridson, in \emph{ACM SIGGRAPH 2007 Sketches} (Association for Computing
  Machinery, New York, NY, USA, 2007), SIGGRAPH '07, p. 22–es.
\newblock \doi{10.1145/1278780.1278807}.
\newblock \urlprefix\url{https://doi.org/10.1145/1278780.1278807}

\bibitem{stomakhian}
A.~Stomakhin, C.~Schroeder, C.~Jiang, L.~Chai, J.~Teran, A.~Selle, ACM Trans.
  Graph. \textbf{33}(4) (2014).
\newblock \doi{10.1145/2601097.2601176}.
\newblock \urlprefix\url{https://doi.org/10.1145/2601097.2601176}

\bibitem{fd27}
D.~Cheng, B.~Chen, X.~Chen, Mathematical Problems in Engineering \textbf{2019},
  13 (2019).
\newblock \urlprefix\url{https://doi.org/10.1155/2019/8532408}

\bibitem{HOULSBY2019103167}
G.~Houlsby, A.~Amorosi, F.~Rollo, Computers and Geotechnics \textbf{115},
  103167 (2019).
\newblock \doi{https://doi.org/10.1016/j.compgeo.2019.103167}.
\newblock
  \urlprefix\url{https://www.sciencedirect.com/science/article/pii/S0266352X19302319}

\bibitem{PhysRevE81011303}
P.W. Humrickhouse, J.P. Sharpe, M.L. Corradini, Phys. Rev. E \textbf{81},
  011303 (2010).
\newblock \doi{10.1103/PhysRevE.81.011303}.
\newblock \urlprefix\url{https://link.aps.org/doi/10.1103/PhysRevE.81.011303}

\bibitem{hutel03199387}
L.~Hu, {Micromechanics of granular materials : Modeling anisotropy by a
  hyperelastic-plastic model}.
\newblock Theses, {Universit{\'e} de Lyon} (2020).
\newblock \urlprefix\url{https://tel.archives-ouvertes.fr/tel-03199387}

\bibitem{Kam15trial}
H.~Askari, K.~Kamrin, Nature Materials \textbf{15}(12), 1274–1279 (2016).
\newblock \doi{10.1038/nmat4727}.
\newblock \urlprefix\url{http://dx.doi.org/10.1038/nmat4727}

\bibitem{contmechnotes}
P.~Papadopoulos, in \emph{Introduction to continuum mechanics} (Department of
  Mechanical Engineering University of California, Berkeley, 2017)

\bibitem{exp1}
N.~Murdoch, B.~Rozitis, S.~Green, T.~Lophem, P.~Michel, W.~Losert, Granular
  Matter \textbf{15}, 129–137 (2013).
\newblock \doi{https://doi.org/10.1007/s10035-013-0395-y}.
\newblock
  \urlprefix\url{https://link.springer.com/article/10.1007/s10035-013-0395-y}

\bibitem{exp2}
N.~Murdoch, B.~Rozitis, S.F. Green, P.~Michel, T.L. de~Lophem, W.~Losert,
  Monthly Notices of the Royal Astronomical Society \textbf{433}(1), 506
  (2013).
\newblock \doi{10.1093/mnras/stt742}.
\newblock \urlprefix\url{https://doi.org/10.1093/mnras/stt742}

\bibitem{exp3}
N.~Murdoch, B.~Rozitis, K.~Nordstrom, S.F. Green, P.~Michel, T.L. de~Lophem,
  W.~Losert, Phys. Rev. Lett. \textbf{110}, 018307 (2013).
\newblock \doi{10.1103/PhysRevLett.110.018307}.
\newblock
  \urlprefix\url{https://link.aps.org/doi/10.1103/PhysRevLett.110.018307}

\bibitem{Fredini2013EvaluationOW}
P.~Fredini, A.~Limache, Comput. Math. Appl. \textbf{66}, 304 (2013)

\bibitem{houdini}
SideFX, https://www.sidefx.com  (2020)

\bibitem{grc1}
H.~Oravec, X.~Zeng, V.~Asnani, Journal of Terramechanics \textbf{47}(6), 361
  (2010).
\newblock \doi{https://doi.org/10.1016/j.jterra.2010.04.006}.
\newblock
  \urlprefix\url{https://www.sciencedirect.com/science/article/pii/S0022489810000388}

\bibitem{kvis}
K.~Skonieczny, S.J. Moreland, V.M. Asnani, C.M. Creager, H.~Inotsume, D.S.
  Wettergreen, Journal of Field Robotics \textbf{31}(5), 820 (2014).
\newblock \doi{https://doi.org/10.1002/rob.21510}.
\newblock
  \urlprefix\url{https://onlinelibrary.wiley.com/doi/abs/10.1002/rob.21510}

\bibitem{refDP}
C.M. Mast, P.~Arduino, P.~Mackenzie-Helnwein, G.R. Miller, Acta Geotechnica
  (2015).
\newblock \doi{10.1007/s11440-014-0309-0}

\bibitem{Nik20eas}
P.~Niksirat, A.~Daca, K.~Skonieczny, The International Journal of Robotics
  Research \textbf{39}(7), 797 (2020).
\newblock \urlprefix\url{https://doi.org/10.1177/0278364920913945}

\bibitem{Kam19trans}
E.R. Parra, K.~Kamrin, Granular Matter \textbf{21} (2019).
\newblock \doi{https://doi.org/10.1007/s10035-019-0948-9}

\end{thebibliography}

\onecolumn
% \listofchanges

\end{document}